\documentclass[lettersize,journal]{IEEEtran}
\usepackage{amsmath,amsfonts}
\usepackage{algorithmic}
\usepackage{array}
\usepackage[caption=false,font=normalsize,labelfont=sf,textfont=sf]{subfig}
\usepackage{textcomp}
\usepackage{stfloats}
\usepackage{url}
\usepackage{verbatim}
\usepackage{graphicx}
\usepackage{amsmath}
\usepackage{makecell}
\usepackage{balance}
\usepackage{tabu}                      
\usepackage{booktabs}                  
\usepackage{lipsum}                    
\usepackage{mwe}                       
\usepackage{multirow}
\usepackage[normalem]{ulem}
\usepackage[hidelinks]{hyperref}
\usepackage{caption}
\usepackage{float}
\usepackage{orcidlink}
\usepackage{algorithm}

\usepackage[most]{tcolorbox}
\usepackage{graphicx}
\usepackage{enumitem}
\usepackage{setspace}
\setlist{noitemsep,parsep=0pt,partopsep=0pt} 

\usepackage{enumerate}
\usepackage{enumitem}
\usepackage{circledsteps}
\usepackage{tikz}

\usepackage[nameinlink,capitalise]{cleveref}
\crefname{figure}{Fig.}{Figs.}
\crefname{table}{Tab.}{Tabs.}
\Crefname{table}{Tbl.}{Tbls.}
\crefname{section}{Sec.}{Secs.}
\Crefname{section}{Section}{Sections}

\usepackage{soul}
\definecolor{black}{HTML}{000000}

\definecolor{dypink}{HTML}{ec008c}
\newcommand{\dyu}[1]{{\color{black} #1}}

\newcommand{\system}{\textit{Havior}}

\definecolor{hbblue}{HTML}{3267B9}
\newcommand{\hb}[1]{{\color{hbblue} #1}}
\usepackage{comment}

\newcommand{\rc}[1]{\textcolor{black}{#1}}
\newcommand{\rw}[1]{\textcolor{black}{#1}}
\newcommand{\tvcg}[1]{\textcolor{black}{#1}}

\usepackage{xspace,xpunctuate}

\newcommand{\eg}{\textit{e.g.},\xspace}
\newcommand{\para}[1]{\vspace{0.6mm}\noindent\textbf{#1}}

\definecolor{cback}{HTML}{EDEFF1}
\definecolor{cframe}{HTML}{B9C4CA}
\newtcbox{\mybox}[1][]{enhanced,
 box align=base,
 nobeforeafter,
 colback=cback,
 colframe=cframe,
 size=small,
 left=0pt,
 right=0pt,
 boxsep=0.8pt,
 #1}

\newcommand{\parasum}[1]{\vspace*{-0cm}\paragraph{\textbf{#1}}}

\usepackage{tcolorbox}
\usepackage{marginnote}
\usepackage{adjustbox}

\global\marginparsep=3pt

\usepackage{mathptmx}                  
\usepackage{CJK}

\AtBeginDocument{

}

\definecolor{review}{HTML}{f72585}

\def\BibTeX{{\rm B\kern-.05em{\sc i\kern-.025em b}\kern-.08em
    T\kern-.1667em\lower.7ex\hbox{E}\kern-.125emX}}
\usepackage{balance}
\begin{document}
\title{Save It for the ``Hot'' Day: An LLM-Empowered Visual Analytics System for Heat Risk Management}
\author{
Haobo Li,
Wong Kam-Kwai,
Yan Luo,
Juntong Chen,
Chengzhong Liu,
Yaxuan Zhang,
Alexis Kai Hon Lau,
Huamin Qu,
and Dongyu Liu

\vspace{-1em}

\thanks{Received 8 May 2025; accepted 3 July 2025. Date of publication 7 July 2025; date of current version 5 September 2025. This work was supported in part by the Hong Kong Research Grants Council under the Areas of Excellence Scheme under Grant AoE/P601/23N, in part by the Theme-based Research Scheme of the Hong Kong Special Administrative Region, China, under Grant T31-603/21-N, and in part by the U.S. National Science Foundation under Grant IIS-2427770. Recommended for acceptance by J. Yang. (Corresponding author: Dongyu Liu.) Haobo Li, Wong Kam-Kwai, Yan Luo, Chengzhong Liu, Yaxuan Zhang, Alexis Kai Hon Lau, and Huamin Qu are with the Hong Kong University of Science and Technology, Clear Water Bay Hong Kong (e-mail: hliem@connec
t.ust.hk; kkwongar@connect.ust.hk; yluodj@connect.ust.hk; chengzhong.liu@connect.ust.hk; yzhangkl@connect.ust.hk; alau@ust.hk; huamin@cse.ust.hk). Juntong Chen is with East China Normal University, Shanghai 200050, China (e-mail: jtchen@stu.ecnu.edu.cn). Dongyu Liu is with the University of California, Davis, Davis, CA 95616 USA (e-mail: dyuliu@ucdavis.edu). This article has supplementary downloadable material available at https://doi.org/10.1109/TVCG.2025.3586689, provided by the authors. Digital Object Identifier 10.1109/TVCG.2025.3586689}}

\markboth{IEEE TRANSACTIONS ON VISUALIZATION AND COMPUTER GRAPHICS,~Vol.~31, No.~10, October~2025}%
{How to Use the IEEEtran \LaTeX \ Templates}

\maketitle

\begin{abstract}
The escalating frequency and intensity of heat-related climate events, particularly heatwaves, emphasize the pressing need for advanced heat risk management strategies. Current approaches, primarily relying on numerical models, face challenges in spatial-temporal resolution and in capturing the dynamic interplay of environmental, social, and behavioral factors affecting heat risks. This has led to difficulties in translating risk assessments into effective mitigation actions.
Recognizing these problems, we introduce a novel approach leveraging the burgeoning capabilities of Large Language Models (LLMs) to extract rich and contextual insights from news reports. We hence propose an LLM-empowered visual analytics system, \system, that integrates the precise, data-driven insights of numerical models with nuanced news report information. This hybrid approach enables a more comprehensive assessment of heat risks and better identification, assessment, and mitigation of heat-related threats. The system incorporates novel visualization designs, such as ``thermoglyph'' and news glyph, enhancing intuitive understanding and analysis of heat risks. The integration of LLM-based techniques also enables advanced information retrieval and semantic knowledge extraction that can be guided by experts' analytics needs. 
\dyu{We conducted an experiment on information extraction, a case study on the 2022 China Heatwave, and an expert survey \& interview collaborated with six domain experts, demonstrating the usefulness of our system in providing in-depth and actionable insights for heat risk management.}
\end{abstract}


\section{Introduction}


Extreme climate events~\cite{machine2023ccarinoa}, particularly those related to heat~\cite{related_weather_reddy_2021}, have seen a marked increase in intensity and frequency in recent years, raising significant concerns globally. 
Heat risks are associated with excess mortality due to temperatures above long-term averages and specific extreme events like heatwaves~\cite{hot2021kristie}.
The adverse impacts extend beyond personal health, as reduced productivity is observed due to heat-related health issues among employees~\cite{increased2021parsons}.
They can further cause severe damage to infrastructure, including buildings, power grids, and roads, leading to disruptions in daily life and economic activities~\cite{exploring2018mathew}.
Developing effective strategies for heat risk management, therefore, becomes increasingly urgent~\cite{risk2013kunreuther}. 

This complex task requires a comprehensive understanding of the interplay between various factors, including meteorological, urbanization, demographic, and socioeconomic factors~\cite{projecting2021yang, hot2021kristie, spatiotemporal2021junyi}.
For example, short-sighted policy solutions like increased reliance on air conditioning during heatwaves, while essential for immediate relief, contribute to climate change and cause surges in energy consumption. This stresses power grids and amplifies the risk of failures~\cite{housing2011maller}.



Numerical models are dominantly used in assessing heat risk by environmental researchers~\cite{heat2010hajat, effects2015yi}.
However, they have notable limitations.
First, their sparse spatial and temporal resolution offers insufficient data-driven support for effective risk management~\cite{downscaling2021yaozhi, evaluating2022mcnicholl}. For instance, the fifth-generation reanalysis (ERA5) data~\cite{era52020hersbach}, one of the most used domain data, yields only one estimate for an area spanning approximately $27.75 \times 27.75 km^2$ per hour. This coarse-grained resolution can only predict normalized results and overlook extreme conditions~\cite{performance2022jin}. 
Second, these approaches aim to predict meteorological variables (\eg temperature and humidity), failing to capture the complex risk dynamics involving human behaviors and social factors~\cite{framework2021nicholas}.
Third, the preparedness and response of society to the risk and the instructions for citizens in numerical models are absent.
News articles about environmental issues complement numerical models by providing detailed descriptions of extreme situations, documenting causes and consequences, and discussing city responses and specific advice for handling heat-related events\cite{what2023michelle}. This information offers retrospective evaluations of management strategies that experts can utilize to refine their approaches. However, efficiently retrieving relevant and valuable news articles remains a significant challenge. The absence of tools for integrating diverse data sources—such as textual news and numerical models—hinders decision-makers from fully leveraging the wealth of information available.

The emergence of large language models (LLMs) offers a new opportunity to address retrieval and integration challenges. LLMs have demonstrated abilities ranging from information extraction~\cite{dagdelen2024structured} to question answering and document retrieval~\cite{evaluating2023nan}. 
\rw{Despite these strengths, integrating environmental news into heat risk management encounters technical challenges: the large volume of news articles challenges LLM token limits, impacting retrieval efficiency; the difficulty of fusing information from varied sources, such as numerical data and textual news, poses a barrier to obtaining a holistic view of heat risks; and the hallucination problem of LLMs (text generated by LLMs that is nonsensical or unfaithful, containing false or misleading information presented as fact) may mislead experts.}

To address those challenges, we develop \system~(Heat Savior), an LLM-empowered visual analytics system that integrates numerical data and textual news catering to the needs of environmental researchers and policy makers involved in heat risk management.
{\system} features a novel ``thermoglyph'' design that utilizes metaphorical representations to enhance experts' comprehension of meteorological conditions.
It supports experts in efficiently retrieving, managing, and navigating a large volume of news articles within a human-in-the-loop retrieval process through hex bin visualizations of topics and news glyphs.
To distill the news' semantic meaning and prevent hallucination, $\system$ further employs LLM techniques including prompt engineering and retrieval augmented generation (RAG)~\cite{patrick2020rag}.
By combining the strengths of numerical models and the rich contexts from news, our system empowers stakeholders to make informed decisions and proactively mitigate heat risks.
Our contributions are:
\begin{itemize}[itemsep=0.5pt,label=$\diamond$]
\item {A novel LLM-empowered pipeline that supports human-in-the-loop retrieval and heterogeneous (numerical and textual) data integration in the context of heat risk management. The pipeline has the potential to be applied to various types of textual documents, not limited to news.}
\item {A visual analytics system, \system, featuring ``thermoglyph'', news glyphs, and visualization of hex bins, allowing experts to explore and visualize heat risk insights interactively. We implement $\system$ through an open-source prototype system\footnote{\url{https://anonymous.4open.science/r/Havior-C58D}}.}
\item {Evaluation through a comprehensive case study, survey, and interview with six experts, showing that valuable insights can be obtained by integrating numerical and textual data using $\system$.}
\end{itemize}

\section{Related Work}
\subsection{Visual Analytics of Climate Risk Data}

Climate risk refers to the potential for uncertain outcomes affecting valuable assets~\cite{emergent2015oppenheimer}, quantified by the probability and impact of hazards like flooding~\cite{related_news_twitter_2016, related_weather_social_2022}, rock cracking~\cite{feng2022evis}, and wildfires~\cite{related_weather_wildfire_2023}.
Visual analytics of climate risk primarily focuses on analyzing \textit{weather simulation data}~\cite{related_weather_survey_2018, related_weather_wrf_2017, related_weather_kappe_2019} to derive insights for mitigating these hazards. 
However, existing tools often lack the necessary granularity to address the spatial-temporal complexity of predictions with adequate explanations~\cite{related_weather_prowis_2024}, making it challenging for stakeholders to capture detailed local variations and fully understand and respond to specific risks.

Several visual analytics systems~\cite{related_weather_airvis_2020, aqeyes2021liu, nayeem2022dcpviz} have emerged to explore the multifaceted spatial-temporal dynamics between meteorological variables. While these systems enhance numerical data analysis, they often neglect human-related factors essential for comprehensive heat risk management~\cite{impact2015jian,related_news_twitter_2016, related_news_evolutions_2020}. Bridging this gap is crucial for better decision-making in climate risk management.

Our approach integrates meteorological data with news to provide a novel perspective on risk analysis. 
This fusion of quantitative simulations with contextual news analysis offers a logical interpretation of quantitative variables and empirical support for subjective narratives~\cite{related_weather_prismatic_2024}.
Hence, our system can navigate users through both “Big” (quantitative) and “Small” (qualitative) data, which are mutually dependent and can enhance each other~\cite{anders2014complementary}.

\subsection{Risk Analysis of Text Data}
Over the past decade, microblog data, notably Tweets, has been used for risk analysis~\cite{sense2011maceachren}, including studies on climate risks~\cite{related_news_twitter_2016, related_news_evolutions_2020}.
Such data offer a direct glimpse into the evolution of public sentiment and suggestions, facilitating rapid reactions to emergencies ~\cite{related_news_twitter_2016,qualitative2017gorro, related_news_evolutions_2020}.
However, the casual format and unstructured nature of microblogs present challenges. They often omit necessary context for analyzing periodic events and can amplify biases~\cite{liu2011recognizing, kwak2010what}, thus limiting their utility for analyzing broader meteorological phenomena.


News articles are more reliable in monitoring natural disasters than microblogs~\cite{related_weather_anthropogenic_2023}.
They provide in-depth descriptions of local conditions, consequences, and underlying reasons behind heat-related incidents.
Authored by professionals, these articles are richer, more coherent, and less noisy in analytical discussions~\cite{what2023michelle} (\eg expert opinions and suggested mitigation strategies), supporting retrospective evaluations of management strategies.
However, the inherent complexity in these documents requires heterogeneous data sources and advanced techniques for effective data structuring and interpretation~\cite{related_news_cohortva_2023}. 
In addition, mapping news to specific climate events has traditionally relied on fuzzy logic~\cite{related_news_zhizhin_2007} due to different detail levels between meteorological data and journalistic reporting.
To connect news with climate events, textual descriptions can align with geospatial visualizations for spatial context.
Direct references and visual cues within the text can enhance user understanding of geospatial context~\cite{related_news_geostories_2021, related_news_kori_2022}.


Inspired by these guidelines, $\system$ highlights the key structured information extracted by LLM and connects them to numerical results through novel visual designs.
These designs maintain a good balance between the precision of numerical models and the semantic richness of textual information.

\subsection{Steering Documents Retrieval}

Retrieving relevant content from large document corpora is a significant challenge in text mining due to the ambiguity of natural language~\cite{related_news_survey_2019}. 
Previous visual analytics systems have proposed solutions based on ranking~\cite{related_news_chae_2012}, similarity~\cite{related_work_trivir_2019}, and removing overlaps~\cite{related_work_similarity_2014}.
However, these methods may overlook low-frequency topics, which are crucial in heat risk management. We introduce a hierarchical hexagon layout that prevents neglecting rare topics while maintaining a tight arrangement and preserving text similarity relations.

LLMs have shown promise in summarizing documents~\cite{stiennon2020learning}, but adapting them to specific domain tasks requires substantial computational resources. Efficient alternatives like prompt engineering are limited by token length constraints~\cite{related_work_promptmagician_2024}, making it challenging to analyze multiple documents simultaneously.
Question-based retrieval systems like DocFlow~\cite{related_work_docflow_2024} enhance search accessibility and efficiency using natural language queries. Inspired by these approaches, we explore using RAG, which integrates external knowledge bases. RAG effectively mitigates common LLM challenges, including hallucinations, reliance on outdated information, and lack of transparent reasoning processes~\cite{related_work_xnli_2023}.
By integrating LLMs for semantic analysis and news retrieval, {\system} enhances machine reasoning and supports human-in-the-loop visual analysis.

\section{Informing the Design} \label{design_study}

To develop a system that effectively supports heat risk management, we conducted a year-long collaboration with two domain experts, \textbf{D(esigner)1} and \textbf{D2}. \textbf{D1} is a professor in environmental science with over thirty years of research experience, and \textbf{D2} is an environmental specialist with more than four years of experience. Additionally, \textbf{D1} is a representative in the `My Climate Risk' scheme, launched by the World Climate Research Programme (WCRP)~\cite{my2021world} to mitigate climate event risks.
Through weekly meetings, we engaged in an iterative design process to identify the specific needs and challenges. 
\rc{Our discussions highlighted several limitations of existing numerical models. While these models provide quantitative analyses, they often fail to capture the complexities of urbanization, demographics, and socioeconomic factors crucial for developing effective heat risk management strategies. The reason is that environmental numerical models often parameterize the physical variables and apply the physical constraints, neglecting to incorporate these non-physical factors into their input and output mechanisms.
Certain insights from these aspects can aid in decision-making processes. For instance, the lack of sky view and green space can exaggerate the impact of high temperatures on mental health patients, leading to increased mortality (found in the first case study, \ref{specific_topic}).
In practice, experts find it challenging to collect comprehensive insights and consequences from these complex factors and integrate them into risk management.
As \textbf{D1} noted, ``\textit{If you ask our numeric model how many people die, it has no answer since it only calculates some meteorological variables.}''
}

Recognizing the potential of news to provide contextual and human-centric insights, we aimed to integrate these textual sources with numerical data. However, integrating heterogeneous data sources posed significant challenges, particularly in creating a comprehensive and user-steerable decision-support pipeline.
Based on the insights from our collaboration, we distilled six design requirements, grouped into three categories: \texttt{Numeric} understanding from climate data, \texttt{Semantic} understanding from news, and \texttt{Integration} of the two.

\para{R1:} \mybox{Numeric} 
\para{Analyze historical and future trends.} \label{R1}
The system should support historical analysis and future forecasting. Experts should be able to load and analyze numerical data of interest at different time points, allowing them to examine past trends and forecast future scenarios effectively.

\para{R2:} \mybox{Numeric} 
\para{Examine spatial meteorological conditions.} \label{R2}
The system should enable the analysis of spatial patterns using familiar visualization forms and analytical methods for domain experts. Features such as spatial zoom in/out and the ability to switch between different variables should be provided to facilitate insights. 

\para{R3:} \mybox{Semantic} 
\para{Support human-in-the-loop news retrieval.} \label{R3}
A multi-step retrieval approach should be developed to retrieve a suitable number of news articles that align with experts' interests. 
\textbf{D1} and \textbf{D2} stated, ``\textit{There are tons of news for me. I need a system to retrieve an appropriate amount of news in different stages of analysis.}''

\para{R4:} \mybox{Semantic} 
\para{Enhance management and navigation among large-scale news.} \label{R4}
With the retrieved news list, efficient management should allow experts to filter and rank news based on numeric and semantic criteria. An easy navigation way among a large amount of news should be facilitated.

\para{R5:} \mybox{Semantic} 
\para{Extract insights from heat-related news.} \label{R5}
The system should possess the abilities of structural information extracting, semantic understanding, and contextual question-answering to help experts gain semantic insights from heat-related news.

\para{R6:} \mybox{Integration} 
\para{Integrate numeric and semantic insights for decision-making.} \label{R6}
The system should integrate the impact of news into the numeric model's results to generate insights.

\section{LLM-Empowered Pipeline}

\begin{figure*}[t]
    \centering
    \includegraphics[width=\linewidth]{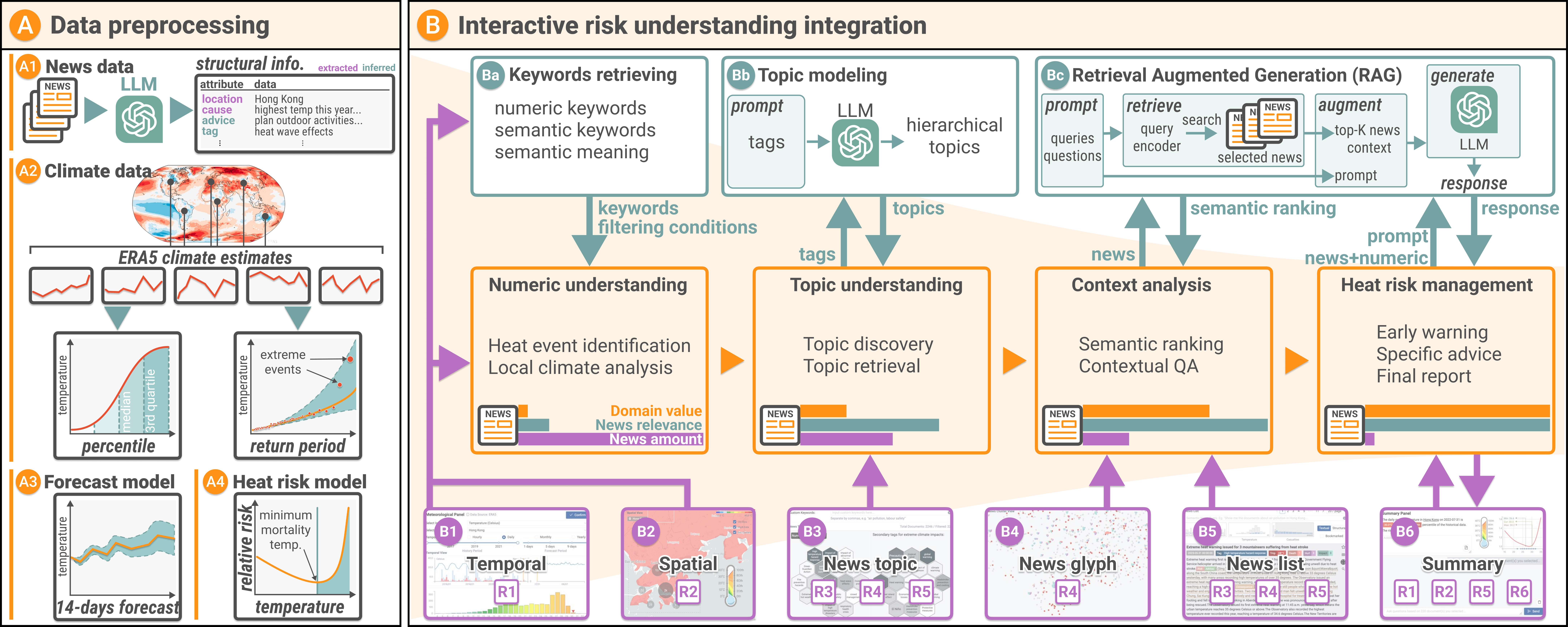}
    \caption{
    The LLM-empowered pipeline contains two parts: data preprocessing (A) and human-in-the-loop risk understanding (B). The data preprocessing involves extracting structural information using LLM (A1) and calculating climate indices (A2-4). In human-in-the-loop risk understanding (B), heterogeneous understandings are integrated through keywords retrieving (Ba), topic modeling (Bb), and RAG (Bc).
    The interactive analysis process is supported by six views of $\system$ (B1-6) which fulfill the design requirements. 
    }
    \label{fig:workflow}
\end{figure*}

\rw{We developed a novel and automated pipeline (\cref{fig:workflow}) that leverages the burgeoning capabilities of LLM to integrate numerical data analysis and semantic understanding to enhance heat risk management.}





\subsection{Data Preprocessing} \label{preprocessing}
\label{sec:data}
\para{Climate data} were obtained from the ERA5 reanalysis dataset~\cite{era52020hersbach}. 
Its utilization and performance in heat risk research have been widely acknowledged for its quality, long-term availability, and accessibility~\cite{mortality2022liu}. 
The hourly data is from 2015 to 2023 and has a spatial resolution of approximately $27.75\ km \times 27.75\ km$. 
While it might seem intuitive to focus on daily temperature extremes (\eg max and min) to assess heat risks, studies have shown that \textit{daily mean temperature} (hereinafter temperature) is more appropriate because it is more comprehensive in representing the day's temperature exposure~\cite{heatwave2018xu}. 
The average of 24-hour (in hours UTC) temperatures within a day was utilized for the temperature for each location.
Moreover, we adopted the Pangu-Weather~\cite{accurate2023bi} model with ERA5 temperature estimates to obtain temperature forecasts up to 14 days (\cref{fig:workflow}-A3).

To contextualize magnitude-based metrics, we also incorporated probability-based indicators for analyzing heat risks.
\begin{itemize}[nolistsep]
    \item 
    \para{Temperature percentiles} offer contexts into the local climatology of a specific region, enabling cross-regional and temporal comparisons~\cite{avoiding2011lass}.  
    This is particularly important when considering variations in local tolerance and preparedness levels~\cite{weather2009anderson}.
    \item \para{Return period} is a statistical measure indicating the estimated average interval between the occurrence of heat events.
    This measure suggests the likelihood of a certain temperature threshold being exceeded at least once a year.
    It evaluates the frequency and intensity of heat events~\cite{risk2013kunreuther}.
\end{itemize}

\para{The heat risk model}~\cite{mortality2015gasparrini} 
examines the relationship between heat-related mortality and temperature fluctuations in 384 locations.
After conducting a sensitivity analysis, the heat-related mortality is quantified using the daily count of deaths for non-external causes only (International Classification of Diseases [ICD]-9 0-799, ICD-10 A00-R99).
The resulting patterns are often represented as ``U'' curves (\hyperref[fig:teaser]{\cref{fig:teaser}}-C1), representing the cold risk and heat risk, with the temperature range associated with the lowest mortality called the Minimum Mortality Temperature (MMT). 
Deviations from the MMT are generally associated with an exponential increase in the relative risk, highlighting the use of temperatures exceeding the MMT when evaluating heat risk.
Since the local climate conditions (\eg tropical or temperate) have a strong influence on individual locations' MMT, our analysis adopts the corresponding heat risk models for each city (\cref{fig:workflow}-A4).

\para{Environmental news dataset} was obtained from Wisers\footnote{\url{https://login.wisers.net/}}, which consisted of 7.7 million environmental (not limited to heat) news articles from Chinese news publishers with over ten years of experience, mainly covering East Asian regions.
The dataset spans from July 2015 to June 2023.
Each news article contains the title, content, character statistics, publishing date, publisher, and media type.
The media types encompass both web and publication resources while excluding internet-based media sources primarily reliant on aggregating news reports from official news agencies.

\rc{
We employed PostgreSQL to store and retrieve news articles.
The computational infrastructure we used is a desktop workstation with an Intel Core i7-12700k processor (3.6 GHz).
For each city, we used SQL queries to conduct the keyword retrieval, specifically selecting news articles containing each target city's name.
Then, we leveraged the GPT-3.5 language model~\cite{language2020brown} to analyze if the news is related to the city's heat risk and extract structural information.
The structural information includes extracted information (\eg location, time, risk description, consequence, reason, temperature, and casualty) and inferred information (\eg advice and tag) (\cref{fig:workflow}-A1).
In cases where details of time are missing, the system utilizes the news article's publication date. This is because news articles prioritize timeliness.
This structural information and the corresponding prompts were developed together with our domain experts---the designers \textbf{D1} and \textbf{D2}. The prompts and examples can be found in the supplement material. 
We utilized the official GPT-3.5 API during the preprocessing.
Due to the large dataset, it takes over two weeks to preprocess all news within one city, including keyword retrieval and structural information extraction.
}

\subsection{Human-in-the-loop Risk Understanding}
\subsubsection{Numerical Climate Data Understanding}
Experts begin by analyzing the numerical climate data to gain a quantitative understanding of the risk (\hyperref[R1]{R1}, \hyperref[R2]{R2}). 
\rw{After the numerical data was preprocessed and displayed in \textit{numeric panel} (\cref{fig:teaser}-A), experts can analyze both magnitude-based and probability-based indices for heat risk and identify heat hazards.}
Additionally, the temporal and spatial information of heat events can help interpret and understand meteorology.
The extracted time and location from news sources are visualized in the \textit{temporal view} (A1) and \textit{spatial view} (A3), respectively. 
Integrating this information with the climate data facilitates experts' numerical understanding of heat risk, which was challenging in the experts' original workflow.

By examining the temporal trend (\cref{fig:workflow}-B1), experts can identify noteworthy patterns and fluctuations in the meteorological variables and the occurrence of news events over time (\hyperref[R1]{R1}). This analysis facilitates the detection of temporal correlations and the identification of significant events or trends within the meteorological context.
Similarly, the spatial distribution of news events (\cref{fig:workflow}-B2) provides valuable information regarding the geographical patterns and localized impacts of meteorological phenomena (\hyperref[R2]{R2}). 
By visualizing the spatial distribution of news articles, experts can discern clusters and patterns of events resulting from heatwaves aiding in the identification of regions affected by specific risks.
Furthermore, we extract numerical temperature and casualties from the news in the preprocessing process. Then, we select the news that documents the highest number of casualties as the representative news for different temperatures (\cref{fig:teaser}-C1), allowing us to provide semantic explanations for numerical risk levels. It transforms the risk level from a mere numerical value into a relatable example, enabling experts to develop a more concrete grasp of heat risks.

\subsubsection{Topic Understanding \& Context Analysis}
To complement the numerical analysis, experts utilize a dataset of environment-related news to gain a semantic understanding of heat risk within the city’s context.

\para{Keywords retrieving (\cref{fig:workflow}-Ba).}
The first step is to retrieve highly relevant news using keywords (\hyperref[R3]{R3}), such as ``Hong Kong,'' ``prolonged,'' and ``high temperature.''
These keywords can be automatically generated based on the numerical analysis results.
For instance, a 97.5th percentile with $\geq 4$ days duration~\cite{heat2017yuming} can lead to the inclusion of the keyword ``heatwave.'' 
They can also be suggested by experts who have analyzed the numerical data. 
\rw{The keyword retrieval is implemented through SQL queries in a PostgreSQL database system.}

\begin{figure*}[t]
  \centering
  \includegraphics[width=\linewidth]{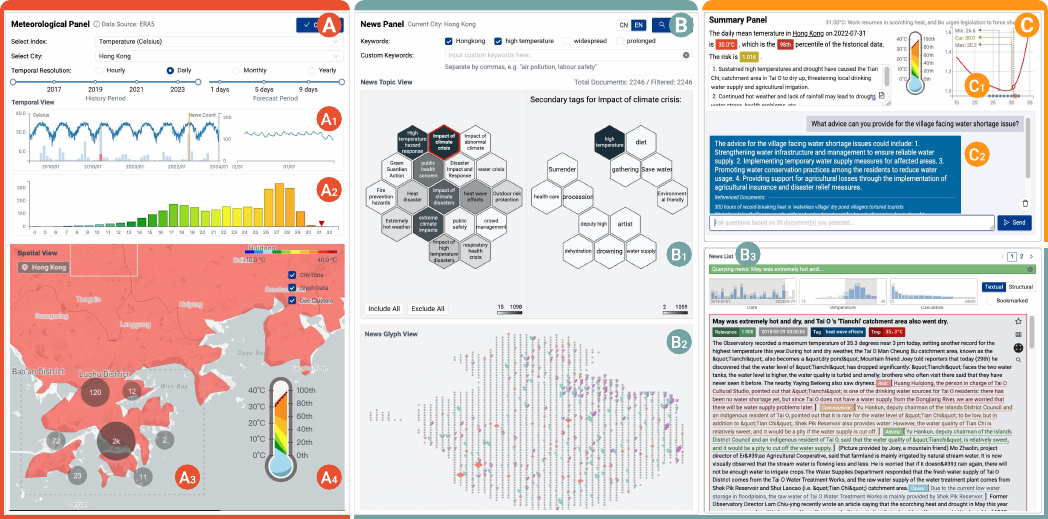}
  \caption{%
    The interface of \textit{Havior} (Heat Savior). 
    The Meteorological Panel (A) facilitates numerical understanding of meteorology, including temporal trends (A1), temporal distribution (A2), and spatial distribution (A3). The ``thermoglyph'' of Hong Kong (A4) intuitively shows the city-based pattern and correlation between temperature and percentile. 
    The News Panel (B) supports human-in-the-loop news retrieval and enhancement in their semantic understanding, in terms of topic-based hierarchies (B1) and risk-based semantic proximity (B2) of retrieved news. The News List (B3) provides details of structural information in the retrieved news with supportive visual cues.
    The Summary Panel (C) enables experts to examine the integration of news and numeric risk model (C1), pose contextual questions (C2), and generate risk management reports.
  }
  \label{fig:teaser}
\end{figure*}

\rc{Then, we employ the GPT-3.5 language model to perform semantic filtering on the retrieved articles, ensuring that only content specifically addressing heat risk (``is heat risk'' is yes) in the target city (``location'' is the target city) is retained.
Additionally, experts can manually apply filters based on rules like ``time,'' ``temperature,'' and ``casualty,'' extracted from the news in the system (\cref{fig:teaser}-B2).
This multi-layered filtering approach ensures both computational efficiency and domain expertise in the final news selection process.}

\para{Topic modeling (\cref{fig:workflow}-Bb).} \label{topic modeling}
\rc{Leveraging the capabilities of LLMs, we extract and cluster tags (a type of structural information, \cref{fig:workflow}-A1) to generate descriptive topics (\hyperref[R3]{R3}, \hyperref[R4]{R4}, \hyperref[R5]{R5}). 
\tvcg{The performance of extraction is evaluated in the \cref{eval_extract}.}
This process creates a two-level hierarchical topic structure. Initially, we extracted three tags per news article to serve as detailed, second-level topics. Subsequently, we apply the DBSCAN~\cite{ester1996density} algorithm to cluster these tags as it is efficient in discovering clusters even with noise and outliers.
Then, the LLM is employed to generate an overarching topic for each cluster, which forms the first-level categorization. The \tvcg{setting of DBSCAN,} the intermediate result of clustering, the prompt for the LLM to generate topics, and the comparison of topic modeling using BERTopic~\cite{grootendorst2022bertopic} and LDA~\cite{blei2003latent} can be found in the supplementary materials. 
}
This approach surpasses traditional named entity recognition and clustering techniques~\cite{huang2024cafellm}, efficiently summarizing and categorizing the retrieved news articles (\cref{fig:workflow}-B3, \cref{fig:workflow}-B4). It serves two objectives: topic discovery and filtering.

First, using LLM, a wide range of comprehensive topics related to heat risk in a specific city can be identified. By exploring these topics, experts can gain an overview of the heat risk landscape and uncover unexpected issues. 
Second, since the number of news articles retrieved about a city often exceeds the experts' capacity to effectively read and analyze, leveraging topics as a criterion enables them to filter the news effectively. Consequently, they can focus their attention on specific areas of interest.

\para{RAG (\cref{fig:workflow}-Bc).}
To mitigate the hallucination problem in LLMs~\cite{shuster2021retrieval}, we employ RAG, which allows experts to pose contextual questions and receive accurate answers by utilizing the retrieved news articles and numerical results as the knowledge source (\cref{fig:workflow}-B6; \hyperref[R5]{R5}).
\rc{Through Embedchain\footnote{\url{https://github.com/embedchain/embedchain}} and PostgreSQL, we integrated the news articles into Embedchain's \textit{app.query()} API, establishing them as a comprehensive reference database.}
This integration broadens the scope of potential questions posed by experts.
To continue enhancing the ability to delve into specific topics, we provide a ranking function based on semantic meaning (\hyperref[R4]{R4}). Therefore, experts can use natural language (sentences or documents) to rank news (\cref{fig:workflow}-B5). 
\rc{This is achieved by designating the retrieved news as the knowledge source and leveraging the semantic meaning to rank the news articles within this source through Embedchain's API, \textit{app.search()}.}
This approach allows news that aligns closely with the experts' interests to be prioritized and placed higher in the ranked list.

\subsubsection{Heat Risk Management}
\tvcg{In the previous analysis process, numeric and semantic content complemented each other to achieve more comprehensive results. For instance, numeric data assisted in retrieving and filtering more relevant news, while the spatiotemporal characteristics and casualty information from the news provided valuable insights to enhance the analysis of numerical data.
In the final stage, }$\system$ integrates two essential perspectives of analysis: numerical and semantic (\hyperref[R6]{R6}) for heat risk management.
The numerical analysis offers quantitative insights into the magnitude, trends, and patterns of heat risk. Its results are utilized for subsequent tasks such as news retrieval, filtering, and comprehension.
The semantic analysis retrieves, filters, clusters, ranks, and analyzes relevant news articles, enabling experts to delve into the context, impacts, and complexities surrounding the risk. 
To summarize, the integration of numerical analysis and semantic understanding allows for a more nuanced assessment, enabling experts to identify potential correlations, causal relationships, and interdependencies among various risk factors.

$\system$ also provides a summary functionality  (\cref{fig:workflow}-B6) that assists experts in consolidating their insights from both numerical and semantic analyses.
Using LLMs, these insights are synthesized into a comprehensive final report, encompassing meteorological conditions, descriptions of heat risk scenarios, historical events or disasters, and advice for government entities and citizens.
This final report serves as a valuable resource for facilitating informed and well-founded decision-making by experts, triggering more effective and rational risk management strategies.
Decision-makers can develop proactive risk mitigation plans, allocate resources effectively, and implement targeted interventions to minimize the negative impacts on economies and the environment.

\section{Visual Design of Havior}
\rc{To develop $\system$, we used modern web technologies. The client is developed using Vite.js, React.js, TypeScript, and D3.js. 
We implement the map using Mapbox GL JS and render visualization elements on the canvas.
We use Zustand to manage application states and handle interactions across different components. The server is implemented using Flask, Flask-Rest-API, and the Postgres database. }

The interface of $\system$ is shown in the \cref{fig:teaser}. To exemplify the connection between views in the interface, let us consider an expert (Zoe) utilizing $\system$ for heat risk research. 
Initially, to check the numerical meteorological condition, Zoe selects the index, city, and temporal resolution from the top menus 
of the \textit{meteorological panel}. The temporal (\hyperref[R1]{R1}) and spatial (\hyperref[R2]{R2}) climate information is then displayed in the \textit{temporal view} and \textit{spatial view}, respectively. 
After gaining the numerical understanding, the next step involves exploring the semantic aspect. Zoe selects the recommended keywords or types the customized keywords (\hyperref[R3]{R3}) in the top menus 
of the \textit{news panel}. 
The \textit{news topic view} displays topics of the retrieved news using hex bins. Zoe can make positive or negative selections of hex bins (\hyperref[R3]{R3}, \hyperref[R4]{R4}), resulting in different scatter plots in \textit{news glyph view} and news lists in the \textit{news list}.
Zoe can easily locate news of interest with the assistance of \textit{news glyph view} and delve into the structural information or full-text of news within the \textit{news list view} (\hyperref[R4]{R4}).

The human-in-the-loop retrieval process and the contextual in the \textit{news panel} question-answering interface in the \textit{summary panel} can help her understand heat-related news (\hyperref[R5]{R5}).
Moreover, insights or knowledge that experts wish to summarize for subsequent reviewing or report generation (\hyperref[R6]{R6}) for decision-making can be pinned to the \textit{summary panel}. 
With the integration of numerical results, $\system$ is able to generate an informative report for decision-making.

\subsection{Temporal View}
The \textit{temporal view} (\hyperref[fig:teaser]{\cref{fig:teaser}}-A1) is for experts to understand the temporal trends and distribution of meteorological variables (\hyperref[R1]{R1}). 
Considering the history period (2015-2023) is typically much longer than the future forecasting period (14 days), line charts for historical and future data are on the same y-axis, but they are separated on the x-axis. 
This separation helps experts distinguish between the known historical data and the projected future data.
To zoom in on a particular timeframe, experts can drag the node of start or end (\includegraphics[width=0.12\textwidth]{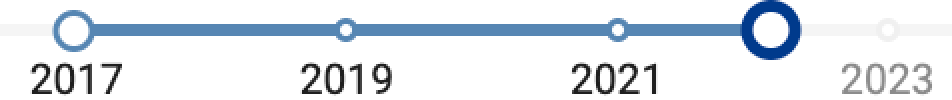}).

Furthermore, they need to explore the temporal relationship between trends and the volume of news related to these parameters (\hyperref[R6]{R6}).
We use the bar chart (\hyperref[fig:teaser]{\cref{fig:teaser}}-A1) to show the number of news articles published.
To save space and easily comparison between the relative magnitudes of changes in both meteorological data and news volume over time, we combine the bar chart and the line chart with the dual y-axis with both y-axes beginning from zero. This design helps in maintaining a visual consistency that can aid in understanding the relationship between the two datasets (\hyperref[fig:teaser]{\cref{fig:rp}}-A, B) without overstating or understating the variations in either due to scaling issues~\cite{study2011isenberg}.
In addition to the temporal trend, we provide a histogram (\hyperref[fig:teaser]{\cref{fig:teaser}}-A2) to display the frequency distribution of temperature (\hyperref[R1]{R1}). 

\subsection{Spatial View}
We provide the \textit{spatial view} aiming to help experts visualize and comprehend the spatial distribution of meteorological variables (\hyperref[R2]{R2}). 
To achieve this, we combine the citywide heat map of the variable (by averaging the data within each city) and the geography map to display the spatial distribution. 
\rc{In determining the color scheme, we draw inspiration from ERA5's color scheme\footnote{\url{https://charts.ecmwf.int/products/medium-2mt-wind30}}, which aligns with domain conventions and is recommended by \textbf{D1-2}.
Compared to the divergent color scheme, which has uniform perceptual changes~\cite{moreland2015we}, it transitions from blue (cold) through green (comfortable) to red (hot), mirroring human thermal perception.}
The spatial relationship between the distribution of meteorological variables and the geographic locations of news articles is vital (\hyperref[R6]{R6}). Thus, we plot news on the map (\hyperref[fig:teaser]{\cref{fig:teaser}}-A3). They will be automatically aggregated in cases of close proximity when zooming in/out, allowing for a more concise visualization. 

\begin{figure}[t]
    \centering
    \includegraphics[width=\linewidth]{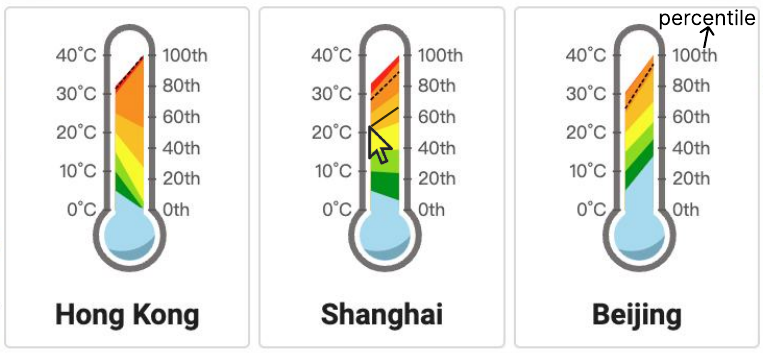}
    \caption{The ``thermoglyph'' in the city gallery for selecting cities. They employ a metaphorical representation. The pattern of color blocks vividly depicts the relationship between temperature and percentile for each city. 
    The black lines connect the current temperature (dashed) or the hovered temperature (bold) to its corresponding percentile.}
    \label{fig:city_glyph}
\end{figure}

\para{Thermoglyph.} 
To effectively visualize both magnitude-based temperature and probability-based index, we have developed the ``thermoglyph.'' They are presented in the city gallery (\cref{fig:city_glyph}), alongside each city on the map (\hyperref[fig:teaser]{\cref{fig:teaser}}-A4), and in the summary panel, which caters to the varying needs of experts throughout different stages of analysis.
The ``thermoglyph'' resembles a thermometer, where the color gradually fills from the bottom to the top, representing the rising mercury in a traditional thermometer due to heat.
The ``thermoglyph'' consists of two parallel axes: temperature and percentile. Different temperature ranges and the associated percentiles are linked and encoded using the same color scheme in the \textit{spatial view}. 
A dashed back line indicates the current value.
We also add a solid black line to illustrate the link accurately when the mouse hovers. 
This feature fulfills the experts' need for precise information.

Consequently, unique patterns emerge in different cities (\cref{fig:city_glyph}). 
\rc{For example, the “thermoglyph” for Hong Kong displays a concentrated pattern on the left side, suggesting that the temperature in Hong Kong is more concentrated for the majority of the time. Due to high exposure, more attention must be given to physiological health, such as cardiovascular disease, as well as mental health~\cite{nazarian2022integrated}.
In contrast, the “thermoglyph” for Beijing illustrates a parallel distribution, indicating a wider temperature range and distinct seasonal characteristics. This presents challenges for the preparedness of citizens, infrastructure, and other factors in response to infrequent high temperatures.
The pattern for Shanghai resembles a state that falls between those of the two aforementioned cities.}
\tvcg{The ``thermoglyph'' of more cities is presented and discussed in the supplementary materials.}

\para{Design alternatives.} 
We consider a design alternative line chart (\cref{fig:workflow}-A2) but find two issues: (1) It is challenging to differentiate between various patterns based on the changing slopes of the lines and (2) caused confusion among users because it implies a temporal change.
Therefore, we opted for the design of the ``thermoglyph,'' which intuitively conveys the relationships between temperatures and percentiles.

\subsection{News Topic View}
\begin{figure}[t]
    \centering
    \includegraphics[width=\linewidth]{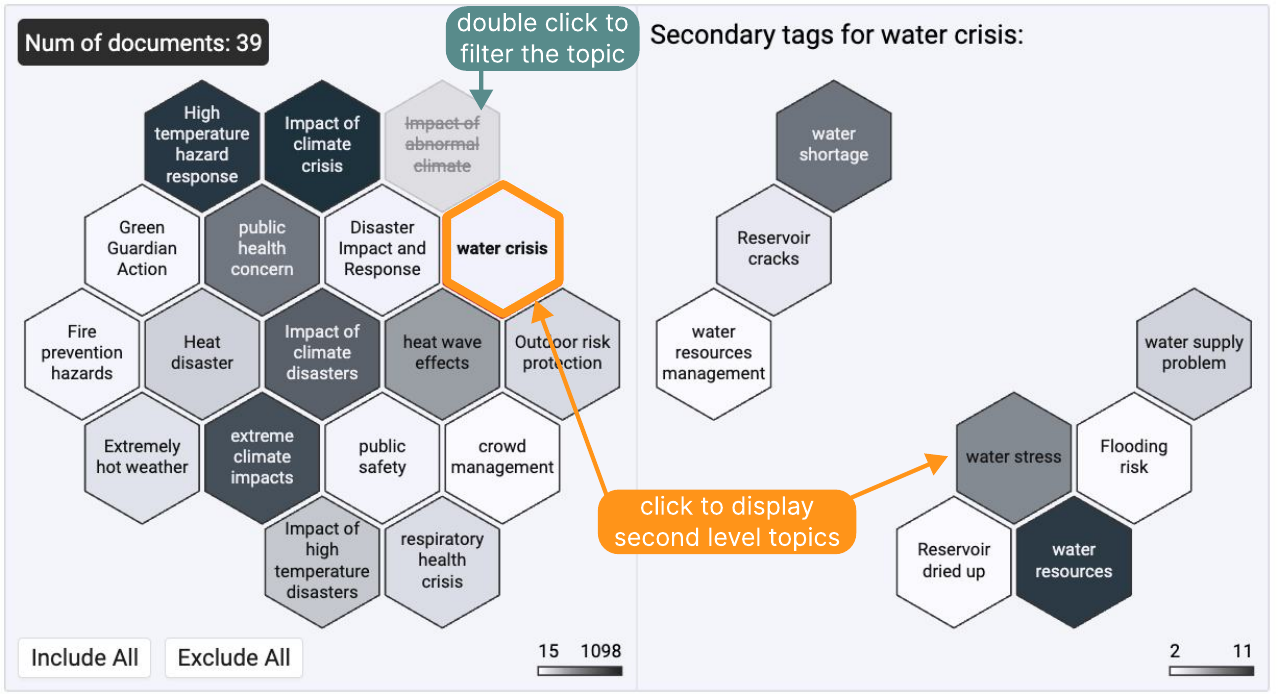}
    \caption{The news topic view displays the hierarchical topics of the retrieved news articles. 
    Left: the first-level topics. Right: the corresponding second-level topics upon clicking a first-level topic.
    Furthermore, double-clicking on a specific topic enables the filtering of news related to that topic in the subsequent analysis process.}
    \label{fig:topic}
\end{figure}

Topic generation of news can help efficiently filter and manage news (\hyperref[R3]{R3}, \hyperref[R4]{R4}) and analysis of news (\hyperref[R5]{R5}). We use a case focusing on the Hong Kong in 2022 China heatwave to illustrate the design of \textit{news topic view} (\cref{fig:topic}).
We employ hex bins with text placed at their centers to represent topics for three reasons. Firstly, compared to \tvcg{a bar chart or} a table list of topics, it has the advantage of conveying the information overview. Secondly, the two-dimensional space of the hex bins preserves the relative spatial relationships between topics~\cite{umap2018mcinnes}, enabling the keeping of semantic meaning relationships. Thirdly, in addition to being used as a container for displaying topics, hex bins are inherently well-suited to serve as buttons for experts to filter news.
We encode the quantity of related news for each topic using a grayscale intensity scheme to avoid confusion with the color in the \textit{spatial view}.

As mentioned in \cref{preprocessing}, we extracted tags for each piece of news.
Then, we cluster these tags and generate the topic for each cluster, resulting in a hierarchical structure.
The first-level topics, which correspond to the cluster titles, provide an overview of the information in each cluster. For example, we know that the ``water crisis'' led to challenges during that particular heatwave in Hong Kong (\cref{fig:topic}, left).
The second-level topics, which are tags of news, offer more specific information pertaining to each cluster. For instance, by clicking the hex bin for ``water crisis,'' additional information such as ``reservoir dried up,'' ``water supply problem,'' and ``water resources'' is revealed (\cref{fig:topic}, right).
The visual design for both first-level and second-level remains consistent.

By directly double-clicking on ``Outdoor work protection'', experts can choose to show or hide the relevant news in the \textit{spatial view}, \textit{news glyph view} and \textit{news list}, enabling them to focus on specific aspects and in-depth analysis.
Furthermore, when the mouse hovers over a topic, the associated bar (\hyperref[fig:teaser]{\cref{fig:teaser}}-A1) in \textit{temporal view} and news glyph (\hyperref[fig:teaser]{\cref{fig:teaser}}-B2) in \textit{news glyph view} will dynamically change color to red.
\tvcg{These designs allow experts to explore the temporal pattern of topics easily.}

\subsection{News Glyph View}

The \textit{news glyph view} (\hyperref[fig:teaser]{\cref{fig:teaser}}-B2) aims to leverage experts' spatial memory for locating and navigating news effectively and enable experts to identify the news of interest quickly (\hyperref[R4]{R4}) with each news represented 
as a glyph in a 2D space.
\tvcg{Suggested by experts (\textbf{D1-2}), news articles reporting higher casualties hold greater importance for them. They also recommend encoding the number of deaths, injuries, and impacted individuals rather than one aggregated number into the glyph. By incorporating this information visually, experts can quickly assess the severity and impact of each piece of news.}
For locating and navigating, they require a reasonable layout of news. 
To achieve this, each news is first encoded as a high-dimensional vector based on its semantic meaning and reduced to two dimensions using UMAP~\cite{umap2018mcinnes}, maintaining semantic similarity. 
\tvcg{The setting of UMAP and }the comparable result of t-SNE can also be found in the supplementary materials.
We also used a grid-based method~\cite{grid2023hilasaca} to avoid clutters and keep the relative distance of news glyphs.
When the topic selection changes, the corresponding news will be shown or hidden without changing the position to keep the consistency of spatial memory.

We opt for dimension reduction and glyph visualization as opposed to ranking visualization methods like lineup~\cite{lineup2013gratzl}, based on several considerations. 
Firstly, news articles encompass a diverse range of topics, and the absence of casualty information does not necessarily diminish their significance (\textbf{D2}). By organizing news articles based on their semantic meaning, we ensure that even those without casualty information are not overlooked entirely. 
Secondly, scalability was taken into account. 
Our glyph design, coupled with a 2D space, provides an effective solution for accommodating a large number of news articles. 
Lastly, the utilization of spatial memory aids in navigation when glyphs maintain interrelationships or distances while retaining intra-meaning through magnitude. The spatial relationships allow for better cognitive mapping and recall of specific articles. Thus, under the condition of a large amount of news, we prioritized an intuitive and space-efficient design. Thus, we opted for a circular form of glyph to represent each news item for both inter and intra benefits.

\begin{figure}[t]
    \centering
    \includegraphics[width=\linewidth]{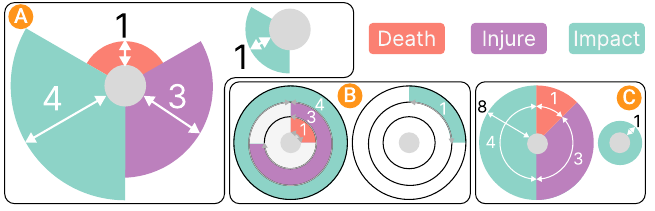}
    \caption{
    We opt for the coxcomb glyph design (A) for the news glyph. Alternative design: target glyph design (B) and pie glyph design (C).
    }
    \label{fig:news_glyph}
\end{figure}

\para{Coxcomb glyph.} 
For circular glyph form, we designed a modified version of the classical coxcomb visualization (\cref{fig:news_glyph}-A) to represent the numbers of deaths, injuries, and impacts. 
To address cases where casualty information is absent, we added a grey node in the center. 
This modification ensures that each news article can be displayed, 
even if the casualty information is unavailable. 
The coxcomb glyph consists of three 120$^{\circ}$ sectors, each distinguished by a different color to represent deaths, injuries, and impacts. The number of casualties within each category is encoded by the length of the sector.
\hb{}
By utilizing the coxcomb glyph, experts can easily identify important news that entails severe consequences.
The news glyph is interconnected with both the \textit{news list} and \textit{spatial view}. Clicking on a glyph will synchronously center and highlight the corresponding news in the other two views.

\para{Alternative designs.} In addition to the coxcomb glyph, we have developed two alternative designs: the target glyph (\cref{fig:news_glyph}-B) and the pie glyph (\cref{fig:news_glyph}-C). However, there are certain limitations associated with them.
The target glyph represents the number of deaths, injuries, and impacts using three individual arcs. It effectively conveys the information.
However, a significant limitation is that it is spatially expensive. Regardless of whether a news article contains casualty information or not, the glyph occupies the same large size.
The pie glyph employs size to encode the sum of deaths, injuries, and impacts. The angle ratio of the sectors is determined based on their respective numerical ratios. 
However, the pie glyph has been criticized for its lack of intuitiveness in expressing individual numbers within each category. Experts need to consider both size and angle to comprehend the magnitude of the numbers.
Considering these factors, we have chosen the coxcomb glyph as the preferred design option. 

\subsection{News List View}
The \textit{news list} (\hyperref[fig:teaser]{\cref{fig:teaser}}-B3) lists the headlines of news articles. 
It provides the functionality for filtering them based on criteria such as time, temperature (which is closely linked to results of domain models), and casualties (\hyperref[R3]{R3}), ranking the news based on their semantic meaning (\hyperref[R4]{R4}), and reading original text or structural information (\hyperref[R5]{R5}).
We plot three bar charts to display the number of news corresponding to each criterion. Then experts can directly brush the bar to apply the filters. The three bars are synchronized. When filtering is applied, a blue color emerges in all bar charts to indicate the number of remaining news.

Experts can use sentences to search and rank news so that they can easily access the news with similar semantic meanings.
To get the details, they can expand the headlines to structural information or full-text. Additionally, visual cues are used to highlight relevant sentences pertaining to ``risk,'' ``cause,'' ``consequence,'' and ``advice'' within full text, which improves the efficiency of reading original text (\cref{fig:teaser}-B3).

\subsection{Summary Panel}
The summary panel was designed to integrate insights of both numeric and semantic to facilitate further decisions to minimize the losses of potential heat risk (\hyperref[R1]{R1}, \hyperref[R2]{R2}, \hyperref[R5]{R5}, \hyperref[R6]{R6}). 
To achieve this integration, we incorporate insights from one side to the other.

\para{Numeric and semantic for numeric.} To provide numerical understanding, the line chart (\hyperref[fig:teaser]{\cref{fig:teaser}}-C1) illustrates relationships between city risks and temperatures. Experts can read the numerical risk level for decision-making.
We also select news with the highest number of deaths at each temperature as representative examples. This approach allows experts to match impacts with different risk levels, contextualizing a mere number.
These representative news are plotted as scatter points on the x-axis, which show a tooltip when hovered.

\para{Semantic and numeric for semantic.} 
Any semantic insights found during the entire process can be pinned to this panel for the purpose of reviewing findings and generating the final report. 
In order to enhance the comprehension of risk, we implement a contextual question-answering interface (\hyperref[fig:teaser]{\cref{fig:teaser}}-C2) that helps experts pose contextual questions. Contextual answers will be generated based on selected news from the \textit{news list} using RAG, which can help with the hallucination problem of LLM.
To address the hallucination problem further from the visual analytic perspective, we link the answer generated by LLM to the source news that LLM references (\hyperref[fig:teaser]{\cref{fig:teaser}}-C2). By clicking the reference, experts can effortlessly access the original news article.
The key results derived from the numerical model are rephrased in natural language for experts to review. They are of utmost importance as they serve as references for generating the final report as well.
By curating and maintaining the content within this panel, a comprehensive final report can be generated using LLM. 
The report becomes a valuable reference for experts working on future actionable plans.

\section{Evaluation}

\rc{\subsection{Evaluation of Extraction}}
\label{eval_extract}

\rc{LLMs have demonstrated remarkable capabilities in extracting structured information from unstructured text~\cite{dagdelen2024structured}. 
To validate the reliability and accuracy in the heat risk context, we designed the experiment.}
\rc{\para{Experiment Setting.}
Due to the lack of ground truth of extraction in this scenario, we implemented a collaborative validation approach involving eight participants across interdisciplinary backgrounds. The participant group comprised four PhD students and one post-doctoral researcher from computer science, alongside two PhD students and one post-doctoral researcher from environmental science. We randomly sampled 50 news articles from our first case study and tasked these participants with evaluating the accuracy of the LLM-extracted structural information.
We developed a user-friendly interface to improve the efficiency of the assessment process (the figure is in the supplement materials). 
The interface was structured to allow participants to simultaneously view the original news on the left panel and provide nuanced judgments using three distinct classification icons: ``Good,'' ``Medium (or Cannot Decide),'' and ``Bad.'' The time duration for each participant was 30-60 minutes.}

\begin{figure}[t]
    \centering
    \includegraphics[width=\linewidth]{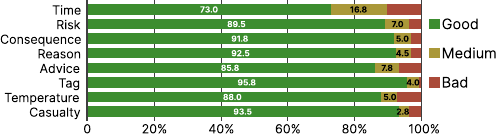}
    \caption{\rc{The result of LLM-based information extraction. Green: the extracted information is accurate. Yellow: medium situation or the participants can not decide. Orange: inaccurate.}}
    \label{fig:extract_eval}
\end{figure}

\rc{\para{Experiment Result.}}
\rc{800 news articles in total are evaluated in terms of information extraction using LLM in the heat risk scenario. The detailed breakdown of extraction accuracy across eight categories is shown in \cref{fig:extract_eval}.
The result demonstrates the good performance of using LLM to extract structural information in the heat risk context.
Specifically, ``Tag'' achieved the highest accuracy at 95.8\%. 
It reveals the summary text power of the LLM. 
``Casualty'' at 93.5\%, ``Reason'' at 92.5\%, and ``Consequence'' at 91.8\% follow closely. All of them exceed 90\%.
``Risk,'' ``Temprature,'' and ``Advice'' extraction also showed strong performance with accuracies of 89.5\%, 88.0\%, and 85.8\%.
}
\rc{While still demonstrating good overall performance at 73.0\%, the most variable category was ``Time''. This variability can be attributed to the inherent challenges in temporal information representation within news articles, where precise temporal details are often incompletely reported. Many news articles tend to omit specific day, month, or year information. To mitigate this limitation, we implemented a robust data completion strategy: when temporal information was incomplete, we utilized the article's publication date to supplement the missing item (day, month, or year), thereby improving the overall extraction accuracy.}

\rc{Overall, these results suggest that our LLM-empowered approach is effective in the context of heat risk management.
}

\subsection{Case Study}
\label{case}
For the case study, we collaborated with experts possessing extensive research experience in environmental science.
\rw{Besides \textbf{D1} (professor, 33 years experience) and \textbf{D2} (specialist, 4 years experience), who we closely collaborate with, we involved four more experts who are not co-authors to conduct the case study. They are \textbf{E3} (professor, 10 years experience), \textbf{E4} (professor, 9 years experience), \textbf{E5} (professor, 10 years experience), and \textbf{E6} (the director of the city's observatory, 30 years experience).
According to Isenberg et al.'s research~\cite{isenberg2013systematic}, collaborators, as well as domain experts, are invited to evaluate the visualization.
Thus, we include \textbf{D1-2}, who are both collaborators and domain experts, in the evaluation to gather their feedback. Their comments are also valuable as they offer suggestions only from the domain-specific viewpoint in the design phase and represent the target users.
}
Regarding our roles~\cite{lantian2022how}, we positioned ourselves as fellow researchers exploring the same problem from different perspectives to facilitate open academic feedback from the professors. For the specialist and the observatory director, we presented ourselves as trustworthy, policy-oriented scholars to elicit practical comments and provide actionable solutions.

The experts studied the 2022 China heatwave~\cite{extreme2023bingqian} in Hong Kong using $\system$. 
The evaluation consisted of two sessions:

\begin{itemize}[nolistsep]
    \item \para{Training and Exploration} (60 minutes): 
    We began with a training session where we demonstrated an interesting case to exemplify $\system$’s functionality and usability. During this period, the experts operated the system under our guidance. Following the demonstration, the experts engaged in a free exploration of the system. 
    \item \para{Survey and Interview} (30 minutes): We then conducted a survey that included a questionnaire and a semi-structured interview to gather ratings and detailed feedback from the experts.
\end{itemize}
%

%

Hong Kong is an exemplary case to delve into the heat risk in light of the increasing occurrence of extreme heat events~\cite{spatiotemporal2021junyi}.

\begin{figure}[t]
    \centering
    \includegraphics[width=\linewidth]{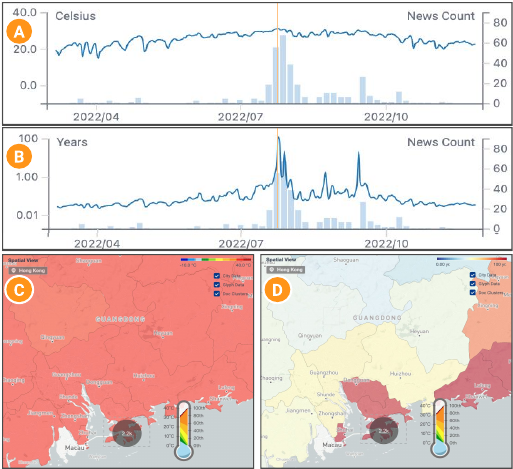}
    \caption{
    The temporal trend (A) and the spatial distribution (C) of the temperature on 2022-07-24 in Hong Kong.
    The temporal trend (B) and the spatial distribution (D) of the return period on 2022-07-24 in Hong Kong.
    Bars in (A) and (B) are the number of news.
    Visualizations of the probability-based return period are better suited for studying extreme heat risk, as compared to magnitude-based temperature. The consistency of heterogeneous results (especially in (B)) enhances the interpretability of the system for heat risk research.
     }
    \label{fig:rp}
\end{figure}

\parasum{Analysis of Numerical Climate Data}
Firstly, the experts choose the index, city and temporal resolution as ``temperature,'' ``Hong Kong'' and ``daily'' to check the meteorological condition. To research the 2022 China heatwave, the experts set the date (single click on \hyperref[fig:teaser]{\cref{fig:teaser}}-A1) to be ``2022-07-24'' since it has the highest temperature (\cref{fig:rp}-A).

The experts commenced their exploration by analyzing the \textit{temporal view} (\hyperref[R1]{R1}) and \textit{spatial view} (\hyperref[R2]{R2}), which effectively presented relevant meteorological information (\textbf{D1-2, E3-6}). \textbf{D2} commented, ``\textit{The geographical representations, color coding, and glyph design provide an intuitive and comprehensive means of understanding of the meteorological condition.}'' 
They discovered that the temperature was anticipated to persist at an alarmingly high level of around 31 degrees Celsius, indicating a prolonged period of extreme heat (\cref{fig:rp}-A). Furthermore, the entire Greater Bay area was engulfed in high temperatures (\cref{fig:rp}-C).
\rw{However, when they examined the ``thermoglyph'' (\hyperref[fig:teaser]{\cref{fig:teaser}}-A4), they found that temperatures between 26–29 degrees are normal in Hong Kong.}
This raised uncertainty about the severity of the situation.

\rc{Realizing the need for a deeper understanding, the experts investigated probability-based indices through the ``Select Index'' dropdown list (\hyperref[fig:teaser]{\cref{fig:teaser}}-A).
Moreover, using the ``thermoglyph'' (\hyperref[fig:teaser]{\cref{fig:teaser}}-A4), they gained a holistic understanding of the relationship between temperature and percentile data (\textbf{D2, E3-4}).}
To their surprise, 
the current percentile for the temperature linked close to the 100th percentile, indicating an unusually severe situation. Seeking further validation, they delved into the detailed information on the return periods.
The large return period (\cref{fig:rp}-B, D) signifies a low probability of the temperature occurrence, which raises concerns about the potentially severe consequences. 
The analysis of return periods provides additional support to the conclusion that Hong Kong is confronted with a substantial heat risk (\textbf{D1, E3, E5}).

\parasum{Analysis of Textual News Data}
The experts turned to the news to enhance their understanding. They conducted a search using the recommended keywords,
resulting in the retrieval of 2,246 news articles.
\rw{By integrating meteorological indices (temperature, return period, percentile) and news articles (\hyperref[R6]{R6}), the researchers found a strong correlation between the return period and the number of news reports (\cref{fig:rp}).
}
The consistency observed between the outcomes derived from heterogeneous climate and news data enhances the rationale for utilizing news with climate data (\textbf{D1, E4-6}).

To obtain an overview of the potential heat risks (\hyperref[R5]{R5}), the experts referred to \textit{news topic view} (\hyperref[fig:teaser]{\cref{fig:teaser}}-B1).
The topics of news are automatically generated by using LLM to cluster tags as mentioned in the \hyperref[topic modeling]{topic modeling} section.
These heat risks encompassed various aspects, including well-known topics such as ``high temperature hazard response,'' and ``impact of climate crisis.'' It was not surprising that the majority of news articles were related to these aspects. However, the experts also encountered unexpected topics, such as the ``water crisis,'' which received relatively less attention.
They found this feature to be highly beneficial in expanding their awareness of previously unrecognized risk topics. They believed that it would enhance their considerations during decision-making processes, leading to more informed and rational decisions (\textbf{D1-2, E5}).
\textbf{E5} commented, ``\textit{The topics found here are not preprogramming, which I think is crucial. When some parameters change or new sources are included, the result can be automatically updated. This is intrinsically natural to pick up new things.}'' 


To observe the temporal pattern of the number of news, the experts analyzed the bar plot showcasing the number of news, which is integrated into the meteorological panel (\hyperref[fig:teaser]{\cref{fig:teaser}}-A1).
Notably, the number of news articles during the summers of 2018 and 2022 stood out significantly compared to other years. This observation aligned with the 2018 southern China heatwave~\cite{record2020deng} and the 2022 China heatwave~\cite{extreme2023bingqian}. Consequently, the experts (\textbf{E4-5}) could infer that a severe heatwave would likely occur if $\system$ were used in 2022.

Motivated by these findings, the experts aimed to explore how Hong Kong responded to heat risks and identify city-based features by deeply delving into specific topics (\hyperref[R5]{R5}). This knowledge would enable the development of more appropriate strategies to assist in preparedness and provide guidance for governments and citizens alike (\textbf{D1-2, E3-6}). 

\begin{figure*}[t]
    \centering
    \includegraphics[width=\linewidth]{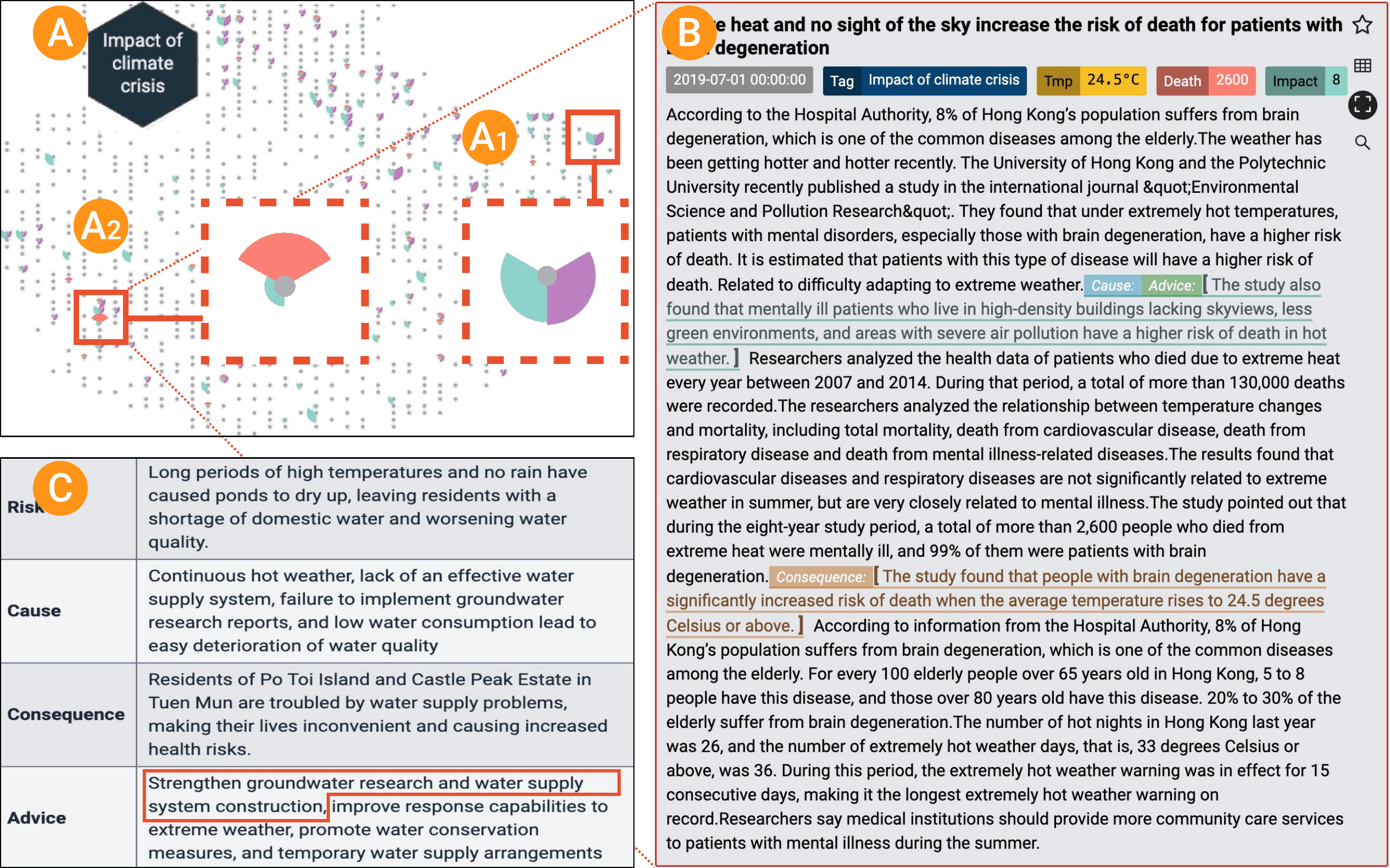}
    \caption{News glyphs under the topic of ``impact of climate crisis'' (A). The largest news glyph (A1) and the news glyph with the largest number of deaths (A2) are likely to be selected.
    The structural information highlighted (B) in the original text helps the experts understand the news quickly. 
    The structural information (C) of the only news with glyph under the topic of ``water crisis'' helps the experts understand the news easily. The advice generated based on summer's news is similar to what the government has taken~\cite{water2018shatian} in 2018 autumn.}
    \label{fig:news}
\end{figure*}

\parasum{Deep Dive into Specific Topics}
\label{specific_topic}
They explored the most well-known topics first by filtering news articles (\hyperref[R3]{R3}) under the categories of ``impact of climate crisis,'' excluding others (\hyperref[fig:teaser]{\cref{fig:teaser}}-B1). 
Then they utilized the news glyph to identify the largest one (\hyperref[fig:teaser]{\cref{fig:news}}-A1) and clicked on it to access the details (\hyperref[R4]{R4}). This particular news discussed the Marathon held in high temperatures that resulted in numerous athlete injuries.
While severe, this outcome was somewhat expected.
The experts bookmarked the news article for future review and pinned important sentences (\hyperref[R4]{R4}) to the summary panel, contributing to the final report.
Subsequently, they came across another news glyph that indicated the highest number of deaths (\hyperref[fig:teaser]{\cref{fig:news}}-A2). 
With the help of visual cues for reading the full text (\hyperref[fig:teaser]{\cref{fig:news}}-B), they can easily gain insights (\hyperref[R5]{R5}).
E3 pointed out, ``\textit{The visualization that maps structural information to the original news text is remarkably neat and useful. You must have dedicated considerable thought to this design.}''.
\rw{They found that this news article highlighted the correlation between high temperatures and increased mortality among mental health patients, particularly in Hong Kong, which lacks sky views and green space. The experts emphasized that such a nuanced correlation would have been difficult to identify solely based on numerical models.}


\parasum{Exploration of Unexpected Topics}
The experts proceeded to investigate the unexpected topic, ``water crisis'' (\hyperref[R5]{R5}). 
They ranked the list of news based on the semantic meaning of the selected news (\hyperref[fig:teaser]{\cref{fig:teaser}}-B3).
By examining the structural information of the first few news in the ranked \textit{news list}, the experts quickly discovered that heatwaves can give rise to water supply challenges, even in a prominent city like Hong Kong.
By referring to the red proportion depicted in the bar (\hyperref[fig:teaser]{\cref{fig:teaser}}-A1), they observed that all news articles concerning tap water supply problems were reported solely during the previous heatwave in summer of 2018 (\hyperref[R6]{R6}).
\textbf{E6} remarked, \textit{``Very good. I remember the drought in 2018 as the observatory director. It should be June, right? (It is June in the system.) The rainfall that year was extremely limited.''}
Noting the government is criticized due to the absence of the tap water supply system for remote villages, the experts posed the question for advice and a considerable answer derived based on the 2018 summer news selected (\hyperref[fig:teaser]{\cref{fig:teaser}}-C2).

Based on the observation that news regarding the unavailability of tap water was only reported during the summer of 2018, the experts inferred that the government had taken measures to tackle the issue.
Our subsequent investigation proved this assumption, as evidenced by a government document~\cite{water2018shatian}.
Additionally, the measures taken by the government in 2018 autumn are very similar to the advice given by $\system$ (\hyperref[fig:teaser]{\cref{fig:teaser}}-C2) and structural information (\cref{fig:news}-C).
This realization also dispels \textbf{E6}'s misconception that the problem been solved due to the rainfall returning to normal after 2018. He is now aware that the government has taken measures as well.
It verifies the ability of $\system$ to discover risk topics and facilitate informed decision-making (\textbf{D1-2, E3-6}).

\parasum{Identifying Unresolved Risks}
On the contrary, some types of disasters were not inadequately handled, such as crop risks due to the heatwave.
\rw{The experts uncovered another unexpected topic regarding ``crop loss'' and ``crop damage.'' These agricultural impacts were attributed to unprecedented insect migration patterns, notably the colonization of Hong Kong by tropical butterfly species, which were fundamentally driven by extreme temperature fluctuations and prolonged drought conditions (\hyperref[R5]{R5}).} These risks were witnessed in 2018, 2021, and recurred in 2022. The lack of attention and proactive preventive measures towards these issues may explain their persistence. However, specific measures are expected to have a positive impact on the risk. Seeking advice from $\system$, they received valuable suggestions such as ``strengthening pest control,'' ``improving the irrigation system,'' and ``using shade nets to reduce plant heat stress.'' With $\system$, they not only identify unresolved heat-related risks but also devise informative strategies to address them (\textbf{D1-2, E3-5}). 

The investigation unveiled an unanticipated phenomenon regarding agricultural vulnerability, specifically documenting instances of significant crop mortality and consequent economic disruption. These agricultural impacts were attributed to unprecedented insect migration patterns, notably the colonization of Hong Kong by tropical butterfly species, which were fundamentally driven by extreme temperature fluctuations and prolonged drought conditions.

\parasum{Summary and Integration of heterogeneous insights}
To integrate the heterogeneous insights for decision-making (\hyperref[R6]{R6}), the experts compiled the semantic insights they discovered in the \textit{summary panel}. Alongside these insights, the numerical conclusions were presented. Additionally, the experts examined the representative news related to the current ($31^{\circ}$) temperature and identified the need for heightened attention to the risks faced by outdoor workers.
By combining the numerical conclusions with the semantic insights from the news, the experts generated a final report on the heat risk in Hong Kong. This report provided them with a comprehensive understanding of the heat risk and helped them to make informed decisions and take appropriate actions (\textbf{D1-2, E4-6}). It can be found in the supplementary material.

\subsection{Expert Survey and Interview}

\rw{To evaluate the effectiveness and usability of $\system$, we developed a questionnaire for the experts based on the goals and requirements of the system.} 
It includes ten 7-point Likert-scale questions (1 = strongly disagree
and 7 = strongly agree) questions and is divided into 2 sections: functionality (assessing usefulness) and usability (evaluating ease of use).
After the case study, the six experts were asked to rate $\system$ from various perspectives using each question.
The result is shown in \cref{tab:result}.

\label{survey}
\parasum{Overall assessment}
All the experts highly rated their experience with $\system$ (all mean scores $\geq$ 6). It is worth noting that the experts are meticulous and cautious in their rating. Particularly, \textbf{E6} thought that for some questions he could not rate immediately and preferred to do so after using the system for more days.
The high score indicates the effectiveness of $\system$, showcasing its ability to manage risks.

\begin{table}[t]
  \setlength{\aboverulesep}{1pt}
\setlength{\belowrulesep}{1pt}
\small
\caption{Ratings on $\system$ on a 7-point Likert scale (with 1 = strongly disagree and 7 =
strongly agree). Questions 1-6 relate to the functionality, and 7-10 relate to the usability of $\system$.}
\label{tab:result}
\makeatletter
\def\hlinewd#1{%
  \noalign{\ifnum0=`}\fi\hrule \@height #1 \futurelet
   \reserved@a\@xhline}
\makeatother
\begin{tabular}{p{0.2cm}| p{6cm} | c | c}
\hlinewd{0.8pt}
\toprule
 & Ratings & Mean & SD \\
\midrule
1 & The meteorological panel helps me understand heat from the quantitative perspective. & 6.33 & 1.21 \\ 
2 & The news panel helps me retrieve the news of interest. & 6.17 & 0.75 \\ 
3 & The news panel helps me manage the news of interest. & 6.00 & 0.89 \\ 
4 & The news panel helps me understand the impact, reasons, and advice of the heat risk. & 6.50 & 0.84 \\ 
5 & The summary panel helps me better understand the risk with the integration of numeric and news. & 6.00 & 0.89 \\ 
6 & The system has the potential to help me make better decisions for heat risk if I am a decision-maker for the city’s policy. & 6.00 & 1.26 \\ 
\midrule
7 & The visual designs in the system are intuitive. & 6.33 & 0.52  \\ 
8 & The interactions in the system are intuitive. & 6.50 & 0.55   \\ 
9 & It was easy to learn the system with the demonstration and training session. & 6.00 & 0.63   \\
10 & I will use this system again. & 7.00 & 0   \\
\bottomrule
\end{tabular}
\end{table}

\parasum{Functionality}
The experts are satisfied with the functionalities offered by $\system$, especially the news panel, which could help them gain different types of insights compared to numerical models, including textual impact and advice (highest mean score and lowest SD). 
One interesting finding is that while the meteorological panel is appreciated by the experts (mean = 6.33), the expert with the longest tenure in environmental research (33 years) assigns the lowest score (score = 4) to the meteorological panel. Despite our efforts to simplify meteorological visualizations, he perceives them as somewhat complex and divergent from his accustomed charts.
It indicates that the experts exhibit ``inertia'' regarding visualization and highlight areas for further improvement. 

\parasum{Usability}
\rw{It was evident that the experts highly valued the visual designs (mean = 6.33) and interactions (mean = 6.50) of $\system$. 
$\system$ only requires the fundamental visualization knowledge (\eg color scheme, shape, trend) and the knowledge of heat risk. Our target users (domain experts) possess the necessary expertise required, making $\system$ accessible to them without extensive prior training.
After the simple training session, they discovered that learning and operating $\system$ was straightforward, evidenced by a mean score of 6.00, indicating a smooth learning curve.
These scores show that our intuitive designs, including visualization and interaction, contribute to a smoother learning curve and enhanced usability. 
As a result, all the experts gave a full score (mean = 7.00, SD = 0) for their intention to use $\system$ in the future.}
The result not only underscores the effectiveness of $\system$, but also indicates a strong inclination towards its adoption for real-world impacts.

\section{Discussion}

\subsection{Implication}
\rc{We reflected on our findings from the entire research process to guide future research directions in visualization and pipeline design for this application and related fields.}

\para{Yesterday's news informs tomorrow's risks.}
The utility of incorporating historical news for dealing with unprecedented extreme events has sparked discussions among both our team and domain experts.
A consensus emerged that while historical insights may have limitations when applied to new or intensely amplified risk scenarios, they can still aid in identifying risks with greater magnitude and tracking unresolved risks.
\textbf{E5} believed that ``there's nothing new under the sun,'' suggesting that new heat extremes still echo those risks in the news.
He expressed that identifying existing risks helps \textit{``extend my scope of consideration by imagining how they scale.}''
The LLM's risk summaries enable experts to anticipate and develop preemptive strategies for more severe extremes.
The recurring risks in the news indicate unresolved problems that need more attention, as in the first case study.
On the other hand, \textbf{D1} praised the system's capability to draw connections between seemingly irrelevant events and reveal new compounded risks.
This analysis, spanning various spatial and temporal dimensions, alerts experts to emerging risks and suggests proactive measures.
\textbf{D1} noted that the significant changes brought by several climate events highlight the potential for risks to converge and intensify, like a chain reaction, eventually creating more significant issues.
He further elaborated, \textit{``These risks have been hidden and cannot be foreseen, but with your system, we now know that they exist and how they evolve.''}

\para{Non-numerical factors in decision-making.}
In the past, the inclusion of non-numerical factors in climate analysis and heat risk management was hindered by the challenge of modeling these unstructured and non-numeric elements, which stand in stark contrast to the domain's usual numerical models.
However, our LLM-empowered pipeline represents significant progress, enabling the seamless incorporation of these factors and their contextualization with the climate model results.
\textbf{E5} expressed, ``\textit{By analyzing highly relevant news, the risk is no longer an abstract number to me. These multifaceted insights help me make informed decisions.}''
Our findings reveal that the extensive documentation (\eg in news and official documents) is now accessible for various critical applications, enhanced by the efficiency of LLMs and visual analytics.

\para{Applicability and Generalizability}
The experts specifically appreciated the vivid ``thermoglyph'' and the valuable insights they obtained through interacting with $\system$.
In particular, \textbf{E4} expressed the satisfaction of effectiveness with $\system$ and complained about the low efficiency and uninspired results of the manual investigation in their original workflow.
\textbf{D1} and \textbf{E6} expected the collaboration of the implementation of $\system$ at the city's observatory to introduce impactful value.
For generalizability, $\system$ provides a feasible pipeline for integrating numeric results and textual insights. The text is not limited to news; other documents, such as government reports, can seamlessly integrate into the pipeline of $\system$.
$\system$ can serve as a source of inspiration for tasks in other domains hurdled by heterogeneous data, especially numeric and textual.

\subsection{Limitation}

\para{The limitation of novel visualization design} is that while new visualizations are beneficial, domain experts may exhibit ``inertia.''
We stroke for simple, yet efficient visualizations recommended by experts during the collaboration.
However, experts' feedback still indicates that the visual design is helpful but requires time to be adapted.
\textbf{E3} commented, ``\textit{I have never seen visualizations like the `thermoglyph' before. However, after a brief introduction, I understand its efficiency.}''
The lesson we learned is the importance of considering the prevalent visualization used in the domain and the underlying reasons. This awareness enables us to design the most suitable visualizations that minimize the transition cost.

\para{The limitations in LLM}, like accuracy, hallucination, robustness, domain expertise, and timeliness, also affect $\system$'s efficiency in providing precise analysis and up-to-date advice.
During the design phase, we took into account the hallucination problem associated with LLM. To enhance the reliability of LLM, we implemented techniques such as RAG, human-in-the-loop analytics, and reference-based Q\&A (\hyperref[fig:teaser]{\cref{fig:teaser}}-C2) to improve the reliability of LLM. As insights are generated during the analytical process and experts have direct access to the source news, the reliability has significantly improved, as noted by experts. 
\textbf{E6} commented, ``\textit{I think the human-in-the-loop exploration process is better than automatically generating a report that I may not trust entirely. It allows me to verify insights in detail by examining the source directly if necessary.}''
However, we still find some inaccurate instances like ``real estate market overheating'' from heat risks.
Our next step is to refine a domain-specific LLM, aiming for a deeper comprehension of heat-related risks.


\para{Visual scalability issue} arises when the news glyph view displays the topic with a large volume of news articles. To address this issue, we have designed a modified coxcomb glyph and implemented a grid-based algorithm to alleviate the problem of visual clutter. In future research, the exploration of more advanced visualization techniques and filtering methods holds the potential to further mitigate this limitation.

\para{Unexplored modalities}, such as crowd-sourced damage photos\footnote{\url{https://www.hko.gov.hk/en/cwsrc/index_mangkhut.html}}, satellite imagery and video footage about disasters suggested by \textbf{E6}, also contain information on risk events.
We seek to incorporate multimodality capabilities in $\system$ for risk management.

\para{\rc{Incorporating more meteorological variables}}
\rc{, such as humidity, could be beneficial. Future research should address the disparity in news coverage, as well as the intricate relationships between these variables and heatwaves.
}

\para{\rc{The uncertainty}}
\rc{of the LLM's results is not directly addressed by the system. Instead, experts interpret the nuances of the LLM's outputs within the human-in-the-loop pipeline, as they can contextualize the findings and their expertise. To enhance this process, uncertainty visualization, such as the variability in the timing or location of news events, can be beneficial. By providing clear visual representations of uncertainty, experts can effectively decide confidence levels.
}

\section{Conclusion}
In this study, we have undertaken the characterization of the risk management problem, with a specific focus on the integration challenges posed by the heterogeneity of numerical results from domain models and risk insights derived from news sources. 
We then developed $\system$, an LLM-empowered VA system, guided by the domain-characterized requirements. $\system$ aims to enhance the analysis of heat risk, improve heat risk management strategies, and mitigate heat-related threats.
The evaluation involved conducting an experiment on information extraction using LLMs and a case study, followed by an expert survey and an interview with six experts. 
Their positive feedback and insights serve as evidence of the usefulness and efficiency of $\system$. Significantly, the evaluation results demonstrate the potential for integrating quantitative model results with heterogeneous insights of risk derived from news reports to enhance the management of heat risks.

\section*{Acknowledgment}

This work was partially supported by the Hong Kong Research Grants Council under the Areas of Excellence Scheme (Grant No. AoE/P-601/23-N) and the Theme-based Research Scheme (Grant No. T31-603/21-N) of the Hong Kong Special Administrative Region, China. It was also supported in part by the U.S. National Science Foundation under Grant No. IIS-2427770.
\bibliographystyle{ieeetr}
\bibliography{main}
\vspace{-16pt}
\begin{IEEEbiography}[{\includegraphics[width=1in,height=1.2in,clip,keepaspectratio]{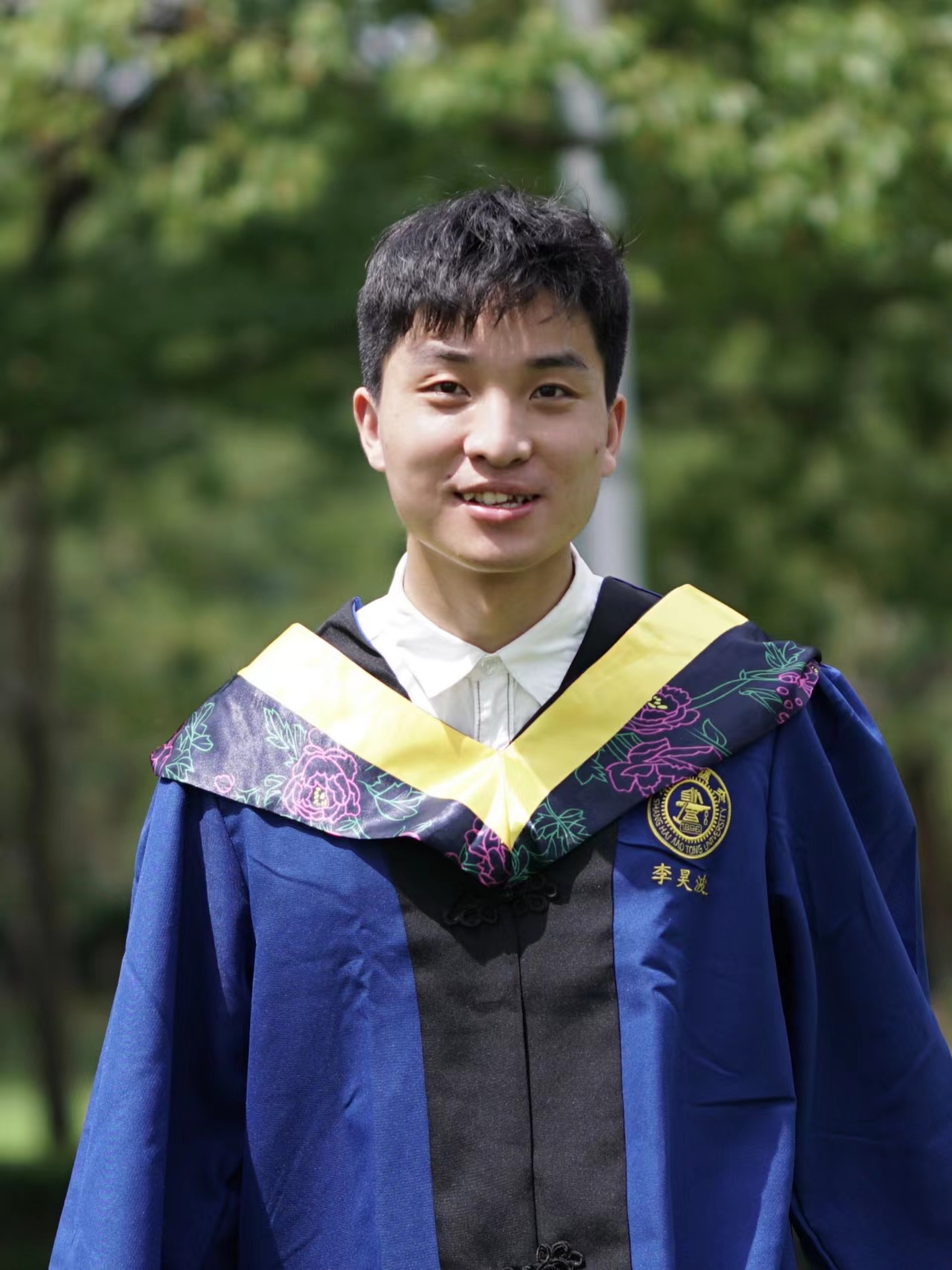}}]{Haobo Li} is a Ph.D. student at HKUST VisLab, Department of Computer Science and Engineering, Hong Kong University of Science and Technology (HKUST). He obtained his bachelor's degree and master's degree from South China University of Technology and Shanghai Jiao Tong University, respectively. His research interests include visualization and the alignment of multimodal climate-related data. For more information, please visit his website at \url{https://hobolee.github.io/}.
\end{IEEEbiography}
\vspace{-12pt}
\begin{IEEEbiography}[{\includegraphics[width=1in,height=1.2in,clip,keepaspectratio]{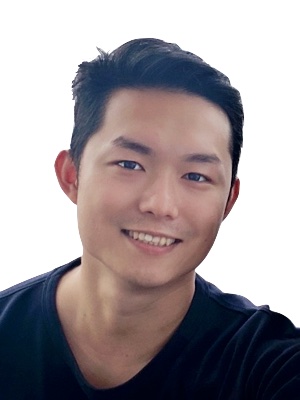}}]{Wong Kam-Kwai} is a Ph.D. candidate in the Department of Computer Science and Engineering at the Hong Kong University of Science and Technology (HKUST). He received his B.E. in HKUST. His main research interests are in data visualization, visual analytics and data mining. For more information, please visit \url{https://kamkwai.com}.
\end{IEEEbiography}
\vspace{-12pt}
\begin{IEEEbiography}[{\includegraphics[width=1in,height=1.2in,clip,keepaspectratio]{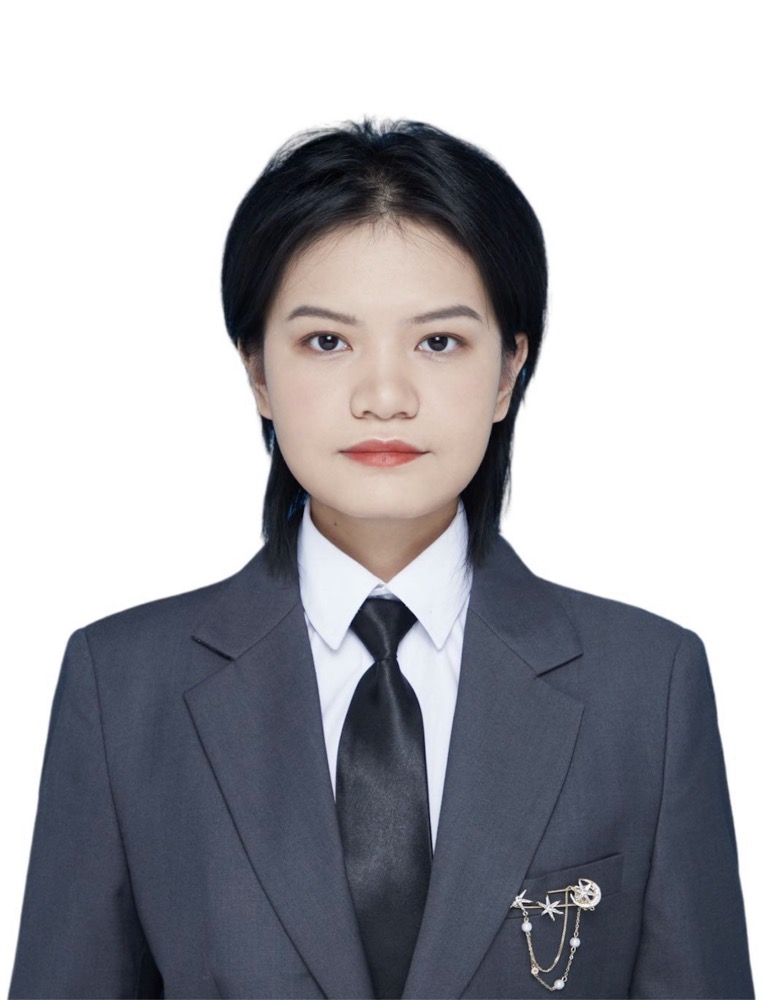}}]{Yan Luo} is currently an MPHIL student in the Department of Computer Science and Engineering at the Hong Kong University of Science and Technology (HKUST). She received her B.E. in HUST. Her main research interests are data visualization, human-computer interaction, and data mining. For more information, please visit \url{https://windyan233.github.io/}.
\end{IEEEbiography}
\vspace{-12pt}
\begin{IEEEbiography}[{\includegraphics[width=1in,height=1.2in,clip,keepaspectratio]{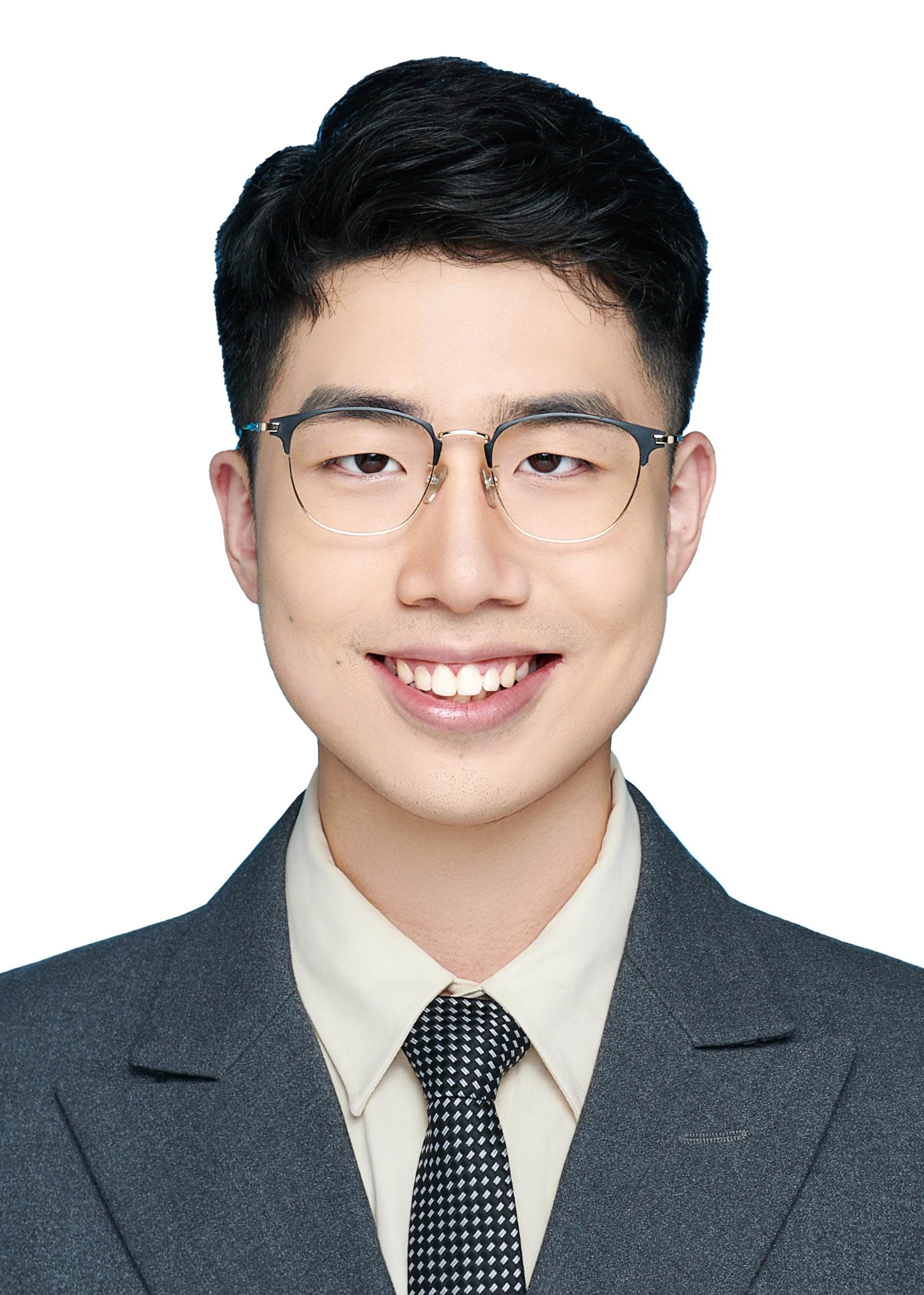}}]{Juntong Chen} received his B.Eng degree from East China Normal University, Shanghai, China, in 2022, where he is currently pursuing his Master’s degree. His research interests include large-scale data visualization, spatiotemporal data analytics, and human-computer interaction, with a focus on urban and climate data. His website is \url{https://billc.io}.
\end{IEEEbiography}
\vspace{-12pt}
\begin{IEEEbiography}[{\includegraphics[width=1in,height=1.2in,clip,keepaspectratio]{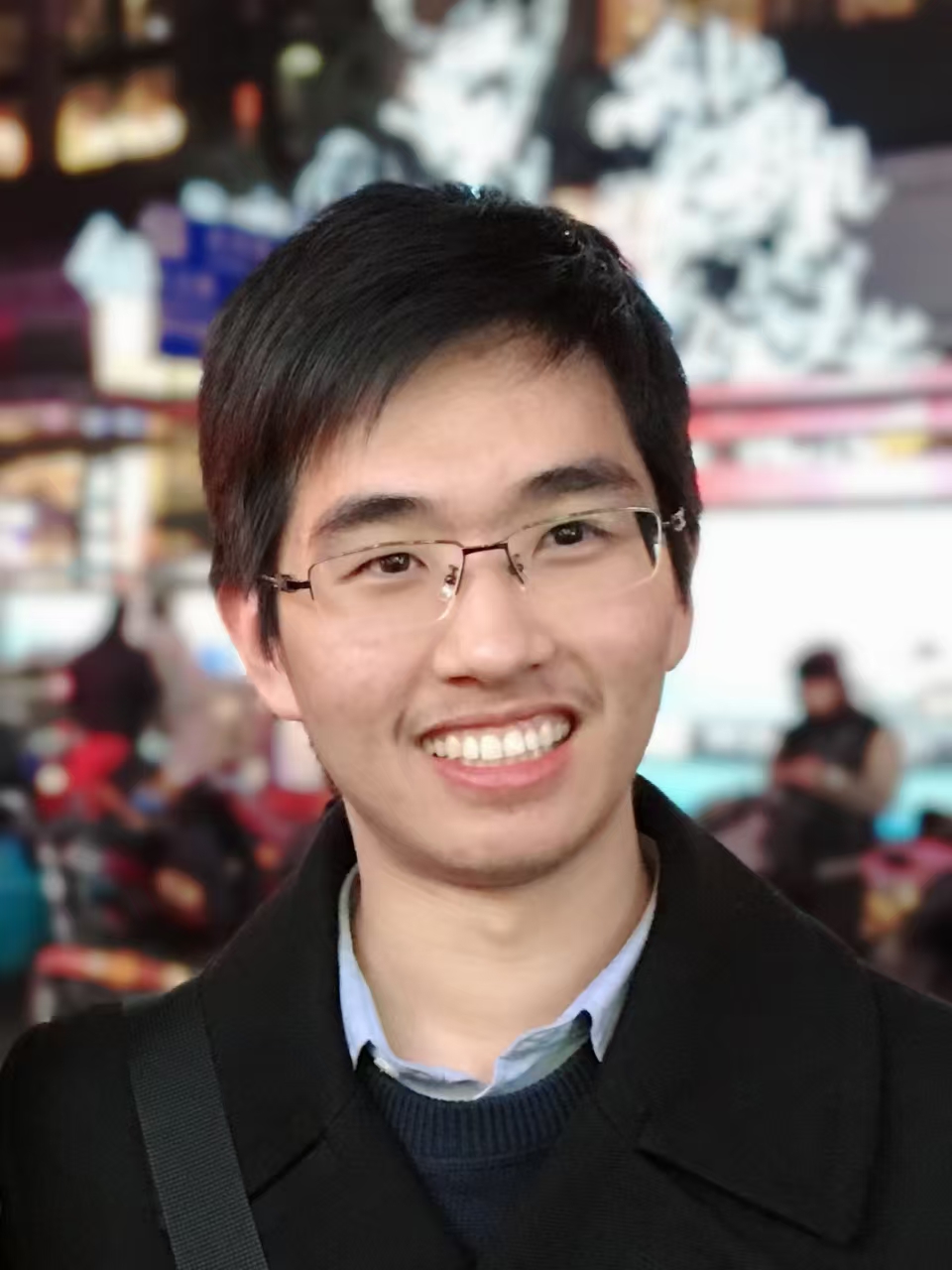}}]{Chengzhong Liu} is a Postdoctoral Research Fellow at HKGAI. He received his Ph.D. in Computer Science and Engineering from the Hong Kong University of Science and Technology (HKUST), supervised by Prof. Xiaojuan Ma. He has published over ten papers in top conferences and journals on HCI and NLP domains, including CHI, CSCW, VIS, and ACL. His research focus is enhancing knowledge-sharing platforms with system design powered by emerging AI tools.  
\end{IEEEbiography}
\vspace{-12pt}
\begin{IEEEbiography}[{\includegraphics[width=1in,height=1.2in,clip,keepaspectratio]{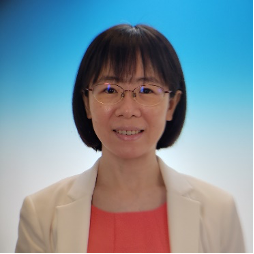}}]{Yaxuan Zhang} is a PhD student in Atmospheric Environmental Science at The Hong Kong University of Science and Technology (HKUST). Her research interests are in Climate Change Early Warning Systems.
\end{IEEEbiography}
\vspace{-12pt}
\begin{IEEEbiography}
[{\includegraphics[width=1in,height=1.2in,clip,keepaspectratio]{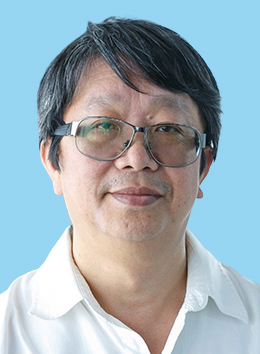}}]{Alexis K H LAU} is the Head of the Division of Environment and Sustainability and a chair professor in the Division of Environment and Sustainability at HKUST. He obtained a PhD from Princeton University. His main research interests are geophysical data analysis, numerical modeling of the atmosphere, regional and urban air pollution, weather and climate, satellite remote sensing applications, and environmental education. More information: \url{https://envr.ust.hk/our-division/people/faculty-staff/alau.html}.
\end{IEEEbiography}
\vspace{-12pt}
\begin{IEEEbiography}[{\includegraphics[width=1in,height=1.2in,clip,keepaspectratio]{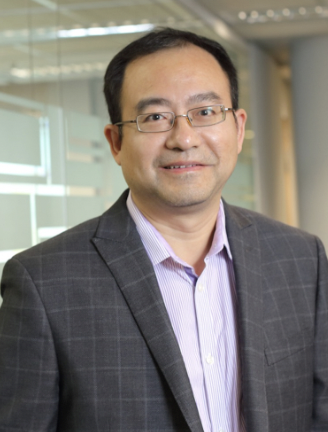}}]{Huamin Qu} is the dean of the Academy of Interdisciplinary Studies, head of the Division of Emerging Interdisciplinary Areas, and a chair professor in the Department of Computer Science and Engineering at HKUST. He obtained a BS in Mathematics from Xi’an Jiaotong University, China, an MS, and a PhD in Computer Science from Stony Brook University. His main research interests are in visualization and human-computer interaction, with focuses on urban informatics, e-learning, and explainable artificial intelligence. More information: \url{http://huamin.org/}.
\end{IEEEbiography}
\vspace{-12pt}
\begin{IEEEbiography}[{\includegraphics[width=1in,height=1.2in,clip,keepaspectratio]{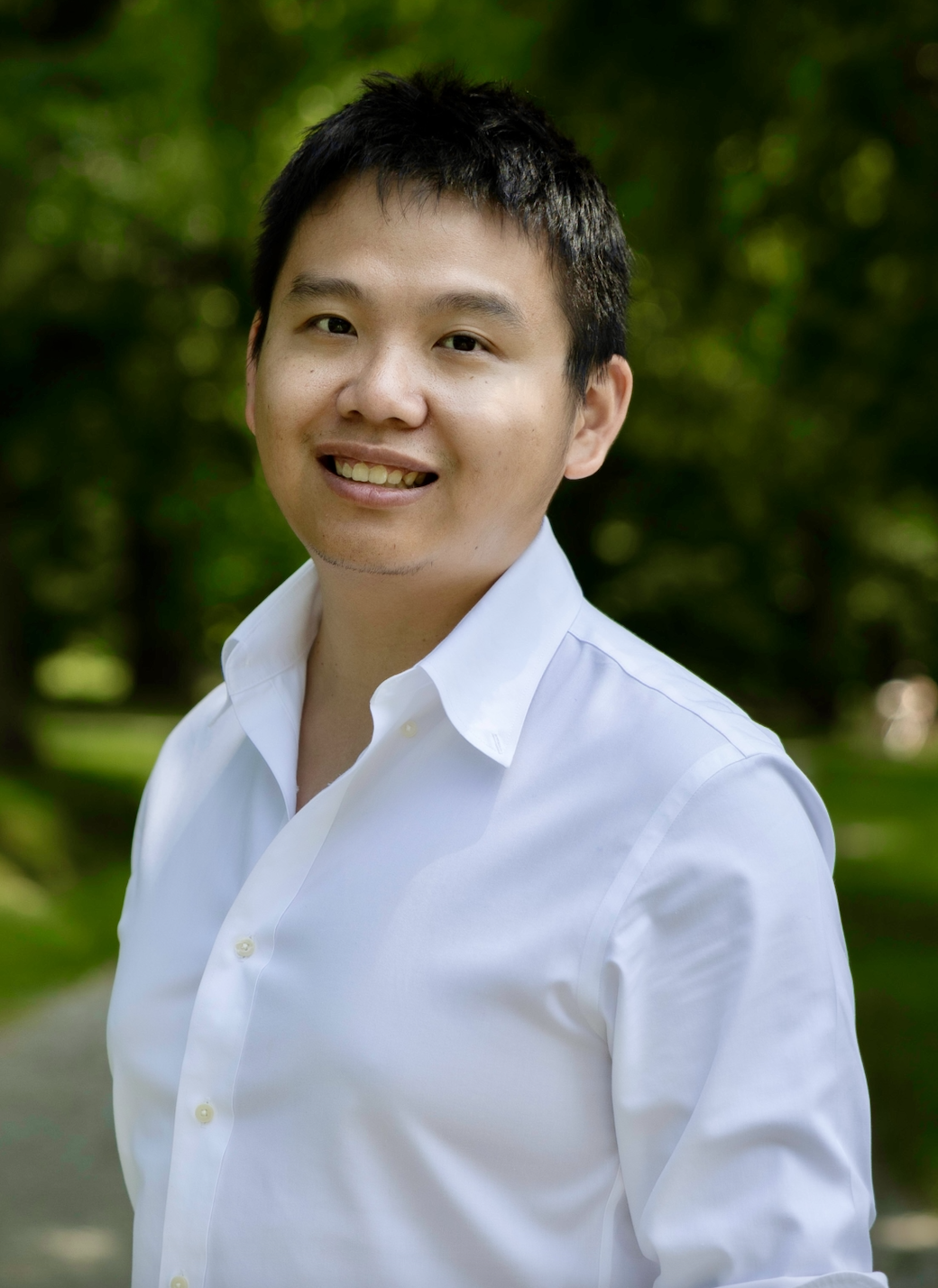}}]{Dongyu Liu} is an assistant professor in the Department of Computer Science at the University of California, Davis, where he directs the Visualization and Intelligence Augmentation (VIA) Lab. His research develops visualization-empowered human-AI teaming systems for decision-making, focusing on sustainability and healthcare. He was a Postdoctoral Associate at MIT and earned his Ph.D. from HKUST. More information: \url{http://dongyu.tech/}.
\end{IEEEbiography}
\clearpage
\end{document}